\documentclass[10pt,times,authoryear,singlespacing,reviewcopy]{elsarticle}
\usepackage{import}

\usepackage[margin=2.5cm]{geometry}

\usepackage[table]{xcolor}
\usepackage{pagecolor}

\definecolor{solarizedLight}{HTML}{fdf6e3}

\color{black}
\usepackage[onehalfspacing]{setspace}  

\usepackage[utf8]{inputenc} 
\usepackage[T1]{fontenc}    
\usepackage[hidelinks]{hyperref}       
\usepackage{url}            
\usepackage{booktabs}       
\usepackage{amsfonts}       
\usepackage{nicefrac}       
\usepackage{microtype}      
\usepackage{lipsum}		

\usepackage{amssymb,amsmath}

\usepackage{subfig,color,mathrsfs,lmodern,bbm}
\usepackage{graphicx}
\usepackage{tabularx}
\usepackage{cases}
\usepackage{colortbl}
\usepackage{natbib}
\usepackage{caption}

\usepackage{url}

\usepackage{algorithm}
\usepackage{algpseudocode}

\algrenewcommand{\algorithmiccomment}[1]{\hskip1em\textit{$//$#1}}

\usepackage{soul,color,xspace,bm}

\newcommand{\be}{\begin{equation}} \newcommand{\ee}{\end{equation}}
\renewcommand{\ee}{\end{equation}}
\newcommand{\bes}{\begin{equation*}}
\newcommand{\ees}{\end{equation*}}

\renewcommand{\emph}{\textit}

\usepackage{placeins}
\journal{arXiv}

\usepackage{xcolor}
\graphicspath{ {./figures_old/} }
\usepackage{amsmath,amssymb}
\usepackage{arydshln} 

\usepackage{algorithm} 
\usepackage{algpseudocode} 
\usepackage{listings} 

\usepackage{multirow}

\usepackage{cases}

\usepackage[automake]{glossaries} 
\usepackage[symbols,nogroupskip,automake,nonumberlist]{glossaries-extra}
\newglossary[slg]{symbols}{syi}{syg}{List of Symbols}
\makeglossaries

\newglossaryentry{temperature}{
    type=symbols,
    name={\ensuremath{T}},
    sort={c1_intro},
    description={temperature [$^\circ$C]; a field of temperature}
}
\newglossaryentry{heat_flux}{
    type=symbols,
    name={\ensuremath{\mathbf q}},
    sort={c1_intro},
    description={heat flux [W/m$^2$]; a field of heat flux}
}
\newglossaryentry{gradient_temperature}{
    type=symbols,
    name={\ensuremath{\mathbf g}},
    sort={c1_intro},
    description={gradient of temperature [$^\circ$C/m]; a field of temperature gradient}
}
\newglossaryentry{thermal_conductivity}{
    type=symbols,
    name={\ensuremath{k}},
    sort={c1_intro},
    description={thermal conductivity [W/(m$\cdot$ $^\circ$C)]}
}
\newglossaryentry{source_term}{
    type=symbols,
    name={\ensuremath{s}},
    sort={c1_intro},
    description={source term [W/m$^3$]}
}
\newglossaryentry{lagrange_lambda}{
    type=symbols,
    name={\ensuremath{\lambda}},
    sort={c1_intro},
    description={field of scalar Lagrange multiplier}
}
\newglossaryentry{lagrange_tau}{
    type=symbols,
    name={\ensuremath{\boldsymbol \tau}},
    sort={c1_intro},
    description={field of vector Lagrange multiplier}
}
\newglossaryentry{boundary_flux}{
    type=symbols,
    name={\ensuremath{\bar q}},
    sort={c2},
    description={function of normal heat flux on boundary $\Gamma_q$}
}
\newglossaryentry{boundary_temperature}{
    type=symbols,
    name={\ensuremath{\bar T}},
    sort={c2},
    description={function of temperature on boundary $\Gamma_T$}
}
\newglossaryentry{order_T}{
    type=symbols,
    name={\ensuremath{p}},
    sort={c2},
    description={approximation order for temperature field}
}
\newglossaryentry{element_size}{
    type=symbols,
    name={\ensuremath{h}},
    sort={c2},
    description={element size [m]}
}
\newglossaryentry{alpha_H1}{
    type=symbols,
    name={\ensuremath{\alpha}},
    sort={c2},
    description={trial function basis indices for $H^1$ space}
}
\newglossaryentry{beta_H1}{
    type=symbols,
    name={\ensuremath{\beta}},
    sort={c2},
    description={test function basis indices for $H^1$ space}
}
\newglossaryentry{gamma_H1}{
    type=symbols,
    name={\ensuremath{\gamma}},
    sort={c2},
    description={change trial function basis indices for $H^1$ space}
}
\newglossaryentry{alpha_L2}{
    type=symbols,
    name={\ensuremath{\mathtt a}},
    sort={c2},
    description={trial function basis indices for $L^2$ space}
}
\newglossaryentry{beta_L2}{
    type=symbols,
    name={\ensuremath{\mathtt b}},
    sort={c2},
    description={test function basis indices for $L^2$ space}
}
\newglossaryentry{alpha_Hdiv}{
    type=symbols,
    name={\ensuremath{\mathcal{A}}},
    sort={c2},
    description={trial function basis indices for $H(\textrm{div})$ space}
}
\newglossaryentry{beta_Hdiv}{
    type=symbols,
    name={\ensuremath{\mathcal{B}}},
    sort={c2},
    description={test function basis indices for $H(\textrm{div})$ space}
}
\newglossaryentry{err_est_order_tolerance}{
    type=symbols,
    name={\ensuremath{t_{p}}},
    sort={c3},
    description={tolerance for the order increase according to the value of average error estimator}
}
\newglossaryentry{err_est_mesh_tolerance}{
    type=symbols,
    name={\ensuremath{t_{h}}},
    sort={c3},
    description={tolerance for the mesh refinement according to the value of average error estimator}
}
\newglossaryentry{err_est}{
    type=symbols,
    name={\ensuremath{\mu}},
    sort={c3},
    description={error estimator per element (face)}
}
\newglossaryentry{noise}{
    type=symbols,
    name={\ensuremath{\eta}},
    sort={c4_DD},
    description={noise applied to the material dataset}
}
\newglossaryentry{noise_std}{
    type=symbols,
    name={\ensuremath{\sigma_{\eta}}},
    sort={c4_DD},
    description={standard deviation of the noise}
}
\newglossaryentry{d_ave_criteria}{
    type=symbols,
    name={\ensuremath{\xi}},
    sort={c5_weak_DD},
    description={comparative value for data-driven refinement}
}
\newglossaryentry{d_ave_tolerance}{
    type=symbols,
    name={\ensuremath{t_{ave}}},
    sort={c5_weak_DD},
    description={tolerance for the average distances to the material dataset}
}
\newglossaryentry{d_std_tolerance}{
    type=symbols,
    name={\ensuremath{t_{std}}},
    sort={c5_weak_DD},
    description={tolerance for the standard deviation of the distances to the material dataset}
}
\newglossaryentry{d_std_ave_tolerance}{
    type=symbols,
    name={\ensuremath{t_{std}^{ave}}},
    sort={c5_weak_DD},
    description={tolerance for the standard deviation of the distances to the material dataset compared to the average distances in element}
}
\newglossaryentry{d_ave}{
    type=symbols,
    name={\ensuremath{d_{ave}^e}},
    sort={c5_weak_DD},
    description={average distance to the material dataset}
}
\newglossaryentry{d_std}{
    type=symbols,
    name={\ensuremath{d_{std}^e}},
    sort={c5_weak_DD},
    description={standard deviation of the distances to the material dataset}
}
\newglossaryentry{perturbation}{
    type=symbols,
    name={\ensuremath{\psi}},
    sort={c6_monte},
    description={perturbation value}
}
\newglossaryentry{std_perturbation}{
    type=symbols,
    name={\ensuremath{\kappa}},
    sort={c6_monte},
    description={standard deviation of the perturbation value}
}
\newglossaryentry{std_temperature}{
    type=symbols,
    name={\ensuremath{\sigma_T}},
    sort={c6_monte},
    description={standard deviation of the temperature result}
}
\newglossaryentry{std_flux}{
    type=symbols,
    name={\ensuremath{\sigma_{\mathbf q}}},
    sort={c6_monte},
    description={standard deviation of the heat flux result}
}
\newglossaryentry{std_grad}{
    type=symbols,
    name={\ensuremath{\sigma_{\mathbf g}}},
    sort={c6_monte},
    description={standard deviation of the gradient of temperature result}
}

\begin{document}

\begin{frontmatter}

\title{Conservative data-driven finite element framework}

\author[uofg]{Adriana Kulikov\'{a}}
\ead{adriana.kulikova@glasgow.ac.uk}

\author[uofg]{Andrei G. Shvarts}
\ead{andrei.shvarts@glasgow.ac.uk}

\author[uofg]{Łukasz Kaczmarczyk~\corref{cor1}}
\ead{lukasz.kaczmarczyk@glasgow.ac.uk}

\author[uofg]{Chris J. Pearce}
\ead{chris.pearce@glasgow.ac.uk}

\affiliation[uofg]{organization={Glasgow Computational Engineering Centre (GCEC), James Watt School of Engineering, University of Glasgow},
            city={Glasgow},
            postcode={G12 8QQ},
            country={UK}}

\cortext[cor1]{Corresponding author}

\begin{abstract}
    This paper presents a new  data-driven finite element framework that is applicable to a broad range of engineering simulation problems.
    In the data-driven approach, the conservation laws and boundary conditions are satisfied by means of the finite element method, while the experimental data is used directly in numerical simulations, avoiding material models.
    Critically, we introduce a ``weaker'' mixed finite element formulation, which relaxes the regularity requirements on the approximation space for the primary field. At the same time, the continuity of the normal flux component is enforced across inner boundaries, allowing the conservation law to be satisfied in the strong sense. 
    The relaxed regularity of the approximation spaces makes it easier to observe how imperfections in the datasets, such as missing or noisy data, result in non-uniqueness of the solution. This can be quantified to predict the uncertainty of the results using methods such as Markov chain Monte Carlo. 
    Furthermore, this formulation provides \textit{a posteriori} error indicators tailored for the data-driven approach, providing confidence in the results and enabling efficient solution schemes via adaptive $hp$-refinement.
    The capabilities of the formulation are demonstrated on an example of the nonlinear heat transfer in nuclear graphite using synthetically generated material datasets.
    This work provides an essential component for numerical frameworks for complex engineering systems such as digital twins.

\end{abstract}




\end{frontmatter}

\section{Introduction} \label{ch:introduction}

Classical computational approaches to engineering problems, such as solid and fluid mechanics, as well as diffusion and transport phenomena, require two components: conservation laws (complemented by appropriate boundary conditions) and constitutive behaviour of the materials. While conservation laws must be obeyed by all systems, any constitutive behaviour is an approximation of reality, often represented by material models obtained by fitting to the experimental data. Material models can vary from simple ones with one or two parameters (e.g. Darcy's flow, linear diffusion/heat flow, isotropic linear elasticity) to more complex ones (e.g. hyperelasticity,  plasticity, unsaturated flow~\citep{vogel_effect_2000_vanGuchten}, nonlinear heat transfer~\citep{matsuo_effect_1980}, among many others).

Fitting material models to experimental data allows for relatively fast and accurate solutions to boundary value problems using physics-based methods, depending on both the complexity of the model and the conformity of the fit. Examples of popular physics-based methods include the finite element method (FEM)~\citep{zienkiewicz_finite_2000}, finite difference method ~\citep{smith1985numerical}, finite volume method ~\citep{versteeg2007introduction}, boundary element method ~\citep{katsikadelis2002boundary}, meshfree methods~\citep{liu2003smoothed} and spectral methods~\citep{boyd2001chebyshev}.
Since the material models are often constructed using phenomenological laws, the existing models can be often reused for materials of similar types, allowing for extrapolation into states not included in the original experiments. For example, applying unsaturated flow models developed for soil~\citep{vogel_effect_2000_vanGuchten} to microfluidic flow in filter paper~\citep{gerlero2022validity}. However, certain materials may not directly fit any established model if the underlying phenomenon is unknown~\citep{singh2022recent}, exhibit changing properties over time due to usage conditions \citep{tzelepi_measuring_graphite_2018, jordan_determining_2018}, or display inherent variance in their responses~\citep{jones2019uncertainty}.
Moreover, the fit of material models to actual material behaviour observed in experiments is often poorly understood, particularly regarding the conditions under which these constitutive relations apply~\citep{edition1986mechanics, belytschko2014nonlinear}. Some materials may also exhibit stochastic behaviour, which is generally averaged out during model fitting, leading subsequent research to seek methods to reincorporate stochasticity into numerical analyses~\citep{rao2013foundations, BENSOUSSAN199151}.

Comparatively to the past, advances in data acquisition have enabled the collection of larger datasets from experimental studies on materials nowadays~\citep{agarwal2014big, baesens2014analytics}. 
The variability of this data, along with the different experimental setups and data collection methodologies, impacts how the data is incorporated into numerical simulations. 
Alternatives to traditional material models include machine learning~\citep{bishop2006pattern}, neural networks~\citep{bishop1995neural}, and deep reinforcement learning \citep{arulkumaran2017deep}, which can either identify material behaviour for future use or find solutions to entire problems. 
However, machine learning approaches which are used on their own may provide solutions that do not adhere to established physical laws~\citep{pateras2023taxonomic}. 
Progress in this research area has led to the development of physics-informed machine learning~\citep{pateras2023taxonomic}, constitutive artificial neural networks~\citep{linka2021constitutive}, and physics-informed neural networks~\citep{raissi2019physics}, bringing the solution back to obeying conservation laws~\citep{grossmann2023physicsinformed}. Nevertheless, machine learning approaches used on their own to obtain a complete numerical solution to mechanical or diffusion problems can be described as a ``black box'', which
can be used without the full understanding of the processes involved~\citep{grossmann2023physicsinformed}. This property, beneficial in some context, can be undesirable in others, where accuracy and control of errors is particularly important, e.g. structural integrity of safety-critical structures. 
To remove the ``black box'' element of the numerical analysis utilising material data, a new approach emerged named data-driven computational mechanics~\citep{kirchdoerfer2016data}

Data-driven computational mechanics provides an alternative to creating material models and allows the use of the data directly in numerical simulations while satisfying conservation laws by means of the finite element method. This approach requires the dataset to define relationships between key variables that would typically be represented in a material model, e.g. stress and strain; heat flux and gradient of temperature; heat flux, temperature and its gradient; etc. These relationships are captured as points within a multidimensional dataset, which can be searched, and where each variable represents a dimension. As data-driven computational mechanics enables the use of material datasets directly instead of constitutive models, this allows the inherent imperfections of the material datasets to influence the results and, therefore, provides means of quantifying uncertainty related to the material behaviour. 

The data-driven approach was initially developed for elastic stress-strain problems and has, since its first introduction, been expanding~\citep{kirchdoerfer2017data, AYENSAJIMENEZ2018752}. 
The applications now include diffusion~\citep{nguyen2020variational}, elasticity~\citep{conti_data-driven_2018}, nonlinear elasticity~\citep{NGUYEN201897}, inelasticity~\citep{EGGERSMANN201981}, plasticity, and fracture~\citep{fracture_data-driven_2020}.  Other research areas also include retrieving material datasets through data-driven identification~\citep{stainier_model-free_2019, leygue2017data, leygue2018data, valdes2022phase}, multiscale analyses using data-driven approaches in one or multiple scales~\citep{mora2020multiscale, karapiperis2021data} and methods for completing incomplete datasets~\citep{ayensa2019unsupervised}. 

Since the original framework of the data-driven computational mechanics can be presented as a formulation with two (or more) independently approximated fields, e.g stress and displacement~\citep{conti_data-driven_2018, korzeniowski_data-driven_2022}, heat flux and temperature~\citep{nguyen2020variational, kulikova_shvarts_kaczmarczyk_pearce_2021}, it can be classified as a multi-field or \textit{mixed} formulation~\citep{boffi2013mixed}. In case of the heat transfer, the original data-driven formulation uses $\mathbf L^2(\Omega, \mathbb{R}^n)$ (Lebesgue) space for the flux and $H^1(\Omega, \mathbb{R})$ (Sobolev) space for the temperature, where $\mathbf L^2(\Omega, \mathbb{R}^n)$ is a space for vector functions with square integrable values and $H^1(\Omega, \mathbb{R})$ is a space for scalar functions with square integrable values and derivatives. This is akin to the choice of spaces in the standard (single-field) finite element approach~\citep{zienkiewicz_finite_2000}, and we will refer to this formulation as the ``stronger'' data-driven formulation.

However, more natural functional spaces can be chosen for each of the fields \citep{brezzi2008mixed, wakeni2022p,boffi2013mixed, arnold1990mixed}. The primary variable in heat diffusion, i.e.  temperature, after integration by parts, can be approximated in $L^2(\Omega)$ space, which allows for the temperature to be discontinuous across the element boundaries. The secondary variable in the mixed formulation is the heat flux, approximated in $H(\text{div}; \Omega)$ space, which ensures the satisfaction of the conservation law in the strong sense and enforces the continuity of the normal flux component across any inner boundaries. A data-driven formulation with the flux field approximated is $H(\text{div}; \Omega)$ space was recently discussed~\citep{rocha2024some}, however, in that work, continuous $H^1(\Omega)$ space approximation of the primary field was used. In this paper, we propose a conservative data-driven formulation using the discontinuous $L^2(\Omega)$ space approximation for the primary field. Following~\cite{demkowicz2024mathematical}, we call this formulation the ``weaker'' DD formulation. 

Although the exact solution for temperature is always in $H^1$ space, $L^2$ space better approximates sharp irregularities. This is particularly relevant for the data-driven method, where the field values taken from the dataset may vary significantly even between different integration points inside one element, for example, in the case of a noisy or incomplete dataset. Furthermore, such choice of spaces in the mixed formulation leads to naturally emerging finite element error indicators and estimators with little computational cost \citep{ainsworth2008posteriori, braess1996posteriori, carstensen1997posteriori}, which can be utilised for adaptive refinements of mesh and/or the approximation order \citep{ZANDER2022103700}. In the proposed framework, the adaptive $hp$-refinement algorithm is guided by the FE error indicators and improved by considering the proximity of the computed fields to the material dataset. This quantity, which were term as the \textit{data-driven} error indicator, permits to ensure that only the elements whose refinement might improve the solution are refined. Such an approach minimises the computational effort required to reach a desired accuracy.

At the same time, using a material dataset instead of a material model in the finite element analysis introduces uncertainty related to the material data into the results. The dataset may be sparse, not cover all the required material states, contain noise or outliers, all of which can impact the simulation accuracy and reliability. In industrial applications, it is important to not only obtain the results of the analysis but also quantify the certainty of these results. In the considered context, the errors/uncertainty of the results may have two sources: the finite element approximation errors and the material dataset imperfections. The FE approximation errors can be controlled by error indicators and reduced using adaptive refinement, ensuring that the effect of the material dataset quality on the uncertainty of the results can be observed.

If the material dataset contains noisy data, the solution of the data-driven approach is not unique and the result can vary depending on the initial field values even when using the same material dataset~\citep{kirchdoerfer2017data}. In such a case, the solution can be perturbed and the analysis repeated to quantify the non-uniqueness of the result. This allows for the quantification of uncertainty, akin to Markov chain Monte Carlo simulations, which provide the average and the standard deviation of the results, informing about parts of the material dataset that are missing points or are noisy. Since this approach requires repeating the simulation multiple times, and the search through the material dataset happens at every integration point, performing the adaptive refinement beforehand is essential, permitting to reduce the number of integration points for a given accuracy and allowing the uncertainty quantification to be computed in an acceptable time frame.

The data-driven framework developed in this paper focuses on nonlinear heat transfer through porous media, such as graphite used in nuclear reactors~\citep{matsuo_effect_1980} or fibrous and foamed insulating materials used in thermal insulation~\citep{larkin_heat_1959}. When dealing with porous media, radiation through the pores might have a significant impact on the thermal conductivity relative to the temperature of the body~\citep{larkin_heat_1959}.
Therefore, the constitutive relationship for porous materials is classically expressed by the Fourier's law where the dependence of thermal conductivity on temperature introduces nonlinearity to the system. To take this into account, in the considered application of the data-driven framework, we utilised a synthetic dataset which includes heat flux, gradient of temperature and the temperature. 

The following section introduces the notation for the diffusion problem used throughout this work, examples of the material dataset and its connection to the finite element formulation. Section~\ref{sec:dd_formulations} contains the derivation of the ``stronger'' (original) and the ``weaker'' data-driven finite element formulations and discusses the differences between them. The verification of both formulations using a problem with a known exact solution is presented in \autoref{sec:verification}. Section~\ref{sec:error_estimates} leverages the functional spaces used by the conservative ``weaker'' DD formulation to adapt the finite element error indicators and additionally introduces new data-driven error indicators, which are used together to form an adaptive $hp$-refinement algorithm. A 2D model of nuclear graphite brick is used for a complete verification for the developed framework for an industrial setting in \autoref{sec:brick}. The graphite brick example continues to be used in \autoref{sec:quantification_non-uniqueness} where the non-uniqueness of the solution is quantified using Markov chain Monte-Carlo analysis to make prediction of the uncertainty of the results. 

All of the numerical analysis in this work has been performed using the specifically developed data-driven module~\citep{kulikova_datadriven_repository} in an open-source parallel finite element library MoFEM\footnote{Website: \href{https://mofem.eng.gla.ac.uk}{\texttt{https://mofem.eng.gla.ac.uk}}.}~\citep{mofemJoss2020}.

\section{Data-driven approach for transport problems} \label{sec:seepage_intro}

This paper focuses on developing a novel, ``weaker'' data-driven approach for heat transport problems. However, the developed formulation can be applied to any scalar transport problem or be extended to vector problems, e.g. elasticity~\citep{kirchdoerfer2016data}.

The conservation of energy for the heat transport problem reads: 
\begin{equation}
    \label{eq:conservation_energy}
    \nabla \cdot \mathbf q = \gls{source_term} \quad \textrm{in} \; \Omega,
\end{equation}
where \gls{heat_flux} $=\gls{heat_flux}(\mathbf x)$ is a field of the heat flux and $s=s(\mathbf x)$ is a field for the heat source in the domain $\Omega$.
The boundary conditions are assumed to be:
\begin{subnumcases}{\label{eq:bcs}}
	\label{eq:boundary_q}
    \mathbf q \cdot \mathbf n = \bar q & on  $\Gamma_q$ \\
	\label{eq:boundary_p}
    T = \bar T & on  $\Gamma_T$,
\end{subnumcases}
where $\bar q$ is the normal flux,  \gls{temperature} $= T(\mathbf x)$ is a field of the temperature and $\bar T$ is its value defined on the corresponding part of the boundary $\Gamma$, see \autoref{fig:potato_bc_general} and \autoref{tab:boundary_conditions_Dirichlet_Neumann}.

\begin{minipage}{0.5\textwidth}
    \centering
    \includegraphics[width=\linewidth]{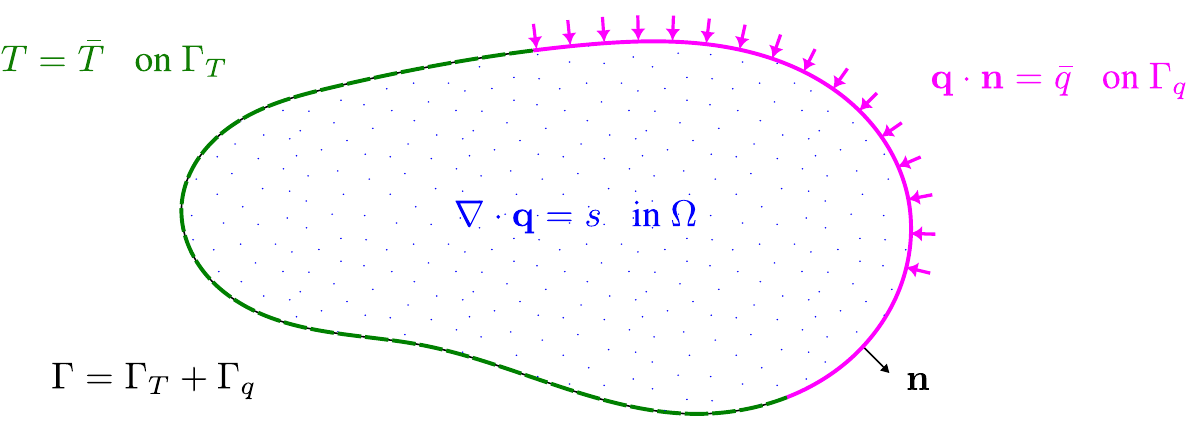}
    \captionof{figure}{Schematic of heat conduction boundary value problem.}
    \label{fig:potato_bc_general}
\end{minipage}\hfill
\begin{minipage}{0.50\textwidth}
    \centering
    \renewcommand{\arraystretch}{1.5} 
    \begin{tabular}{ >{\centering\arraybackslash}p{0.18\textwidth} 
        | >{\centering\arraybackslash}p{0.4\textwidth} : >{\centering\arraybackslash}p{0.2\textwidth}}
        BC type & Field on which the condition is applied & Boundary \\
        \hline    
        Dirichlet & Primary variable: temperature $T$ & $\Gamma_T$ \\
        \hdashline
        Neumann & Secondary variable: heat flux $\mathbf q$ & $\Gamma_q$ \\
    \end{tabular}
    \captionof{table}{Types of boundary conditions for the problem.}
    \label{tab:boundary_conditions_Dirichlet_Neumann}
\end{minipage}
\\

To be able to solve a heat diffusion boundary value problem \eqref{eq:conservation_energy}-\eqref{eq:bcs}, in the standard FE approach, the flux $\mathbf q$ is defined with the help of a phenomenological constitutive equation, i.e. Fourier's law: $\mathbf q = -k \nabla T$. 
However, thermal conductivity \gls{thermal_conductivity} can introduce nonlinearity to the system by the dependence on the unknown temperature $T$, i.e. $k=k(T)$. 
Additionally, the thermal conductivity can be dependent on the spatial heterogeneity, such as the porosity of the material, which may also evolve in time. The resulting constitutive relation could be highly nonlinear and depend on a set of empirical parameters~\citep{ahtt5e}. 

On the contrary, in this paper, following~\cite{kirchdoerfer2016data}, we assume no constitutive equation, while the necessary relationship between the heat flux $\mathbf q$ and the gradient of temperature $\nabla T$ is introduced through a material dataset.

To replace the constitutive equation, the dataset dimensions need to contain the temperature gradient $\mathbf g^*$ and the heat flux $\mathbf q^*$. The data-driven approach consists of two main parts:
\begin{enumerate}
    \item Searching through a material dataset $\mathcal D$ to find a subset of the closest material points to the current solution fields.
    \item Solving the finite element problem using that subset of the material dataset.
\end{enumerate}

To be able to visualise the material dataset, and hence the search through it, we begin with an introduction of material datasets and their handling, and explain the connection to the finite element formulation afterwards.

\subsection{Material datasets} \label{sec:dataset_creation}

\begin{figure}  [tb]
	\centering
    \includegraphics[width=\linewidth]{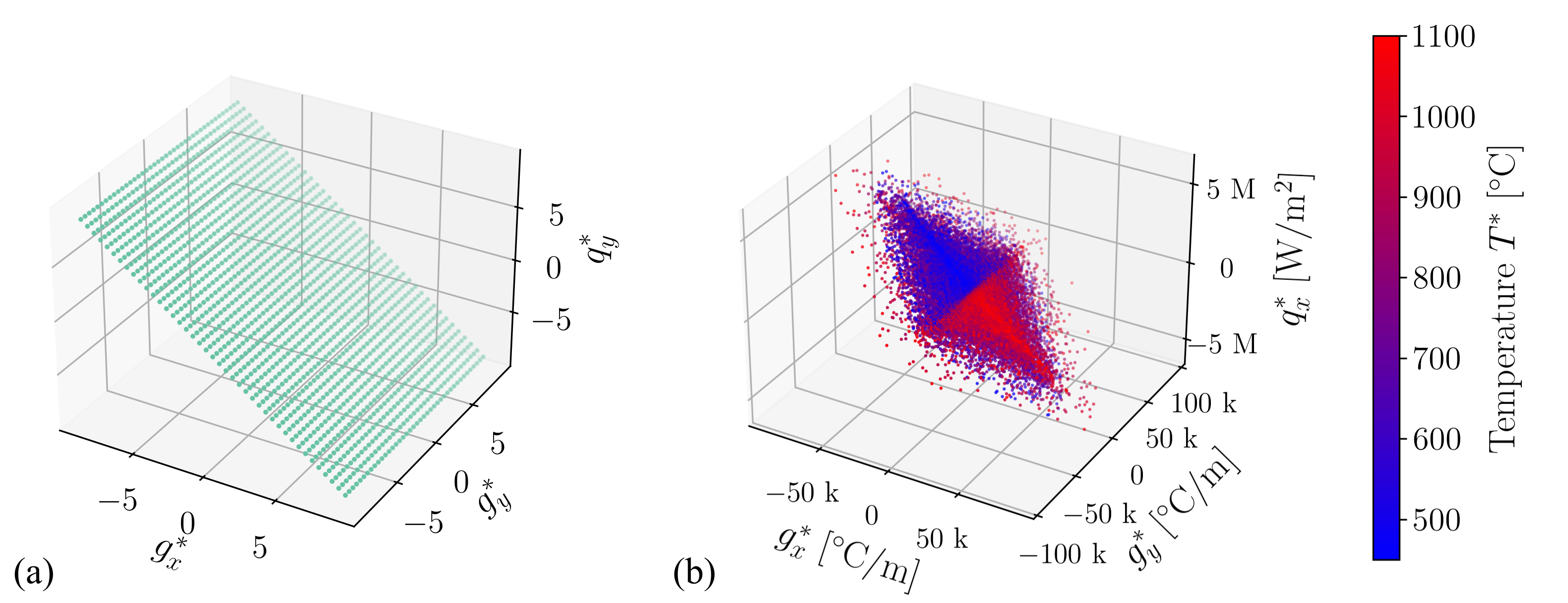}
	\caption{Example material datasets (a) $\mathcal{D}_{4D} = \{g_x^*, g_y^*, q_x^*, q_y^*\}$ and (b) $\mathcal{D}_{5D} = \{T^*, g_x^*, g_y^*, q_x^*, q_y^*\}$. The first dataset is a regular grid and the second dataset is generated through artificial experiments with a nonlinear thermal conductivity of graphite.
     The datasets' creation is explained in \ref{app:dataset_creation}.}
	\label{fig:artificial_datasets}
\end{figure}

For this paper, all of the material datasets are generated for a 2D problem.
Any material points originating from the dataset will be denoted as $(\cdot)^*$, e.g. $\mathbf g^*$ represents the temperature gradient values from the material dataset. For demonstration and verification purposes, two artificially generated material datasets are used in this paper. The datasets are visualised in \autoref{fig:artificial_datasets} and their creation is explained in \ref{app:dataset_creation}. 

The first material dataset consists of points with four variables $\{g_x^*, g_y^*, q_x^*, q_y^*\}_{4D}$, which are spaced evenly in a grid structure, see~\ref{app:regular_dataset} for more details. The grid-like datasets could be obtained by running an analysis akin to representative volume element (RVE) analysis used in multiscale modelling to create the material dataset, which is then used in the macroscopic analysis in place of a fitted model, if the material behaviour is well characterised at the microscale\citep{HUANG2021113013,korzeniowski2022data}.  

The second material dataset is generated through artificial experiments with a nonlinear constitutive law for graphite~\citep{mceligot2016thermal}:
\begin{equation}
 \label{eq:graphite_material_model}
 k(T) = 134 - 0.1047 T + 3.719 \times 10^{-5} T^2,
\end{equation}
which results in five variables $\{T^*, g_x^*, g_y^*, q_x^*, q_y^*\}_{5D}$, where $T^*$ is the temperature, see~\ref{sec:artificial_experiment}. This model is derived from the research on high-temperature gas-cooled reactors, encompassing temperature ranges from room temperature to $1000 ^\circ$C. The dataset obtained from the artificial experiments could be compared to a dataset obtained by data-driven identification~\citep{stainier_model-free_2019, leygue2017data, leygue2018data, valdes2022phase}, where only a part of the data originates from the experiment. 

It is important to note that additional variables, such as porosity or humidity, can be included in the material dataset according to an assumed dependency and the availability of the material data.

\subsection{Search for the closest point in the material dataset} \label{sec:dataset_search}

Assuming a 5D material dataset $\mathcal D_{5D} = \{T^*, g_x^*, g_y^*, q_x^*, q_y^*\}$, the search through it can be performed by minimising the distance of the unknown fields to the dataset. For every point $(x,y)$ in the domain $\Omega$, we can define the distance of the values of the temperature $T$, temperature gradient $\nabla T$ and heat flux $\mathbf q$ to the dataset $\mathcal{D}_{5D}$ as follows:
\begin{equation}
    \label{eq:distance_dataset_dd}
    \textrm{dist} \left(\{T, \nabla T , \mathbf{q} \}, \mathcal{D}_{5D}\right) = \min_{\{T^*, \mathbf{g^*}, \mathbf{q^*} \} \in \mathcal{D}_{5D}} \sqrt{S_{\textrm T}\left( T -T^*\right)^2 + S_{\textrm g}\left(\nabla T -\mathbf g^*\right)^2 + S_{\textrm q}\left(\mathbf q - \mathbf q^*\right)^2} ,
\end{equation}
where $S_{\textrm T}$, $S_{\textrm g}$ and $S_{\textrm q}$ are stabilisation/scaling parameters that depend on a numerical scheme and, in our case, make the distance between dataset points unitless. The material dataset and the dataset search can include more or less dimensions than the five presented above, e.g. the temperature $T$ can be omitted as is the case with the $\mathcal{D}_{4D}$ dataset shown in \autoref{fig:artificial_datasets}(a). However, the temperature gradient $\mathbf g^*$ and heat flux $\mathbf q^*$ always need to be included in the material dataset and in the dataset search.  

The implementation in this paper uses R-tree index structure for spatial searching \citep{guttman1984r}. However, the minimisation of the distance in~\eqref{eq:distance_dataset_dd} can be achieved by other algorithms, such as approximate nearest-neighbour, $k$-means tree, geometric near-neighbour access tree or algorithms based on Mahalanobis distance, see \citep{eggersmann2021efficient, bahmani2021kd, AYENSAJIMENEZ2018752}.

\subsection{Connection of material data to finite element formulation} \label{sec:dataset_fem}

Eq.~\eqref{eq:distance_dataset_dd} can be used to find the closest material data points to the current solution fields at any point in the domain.
A functional representing the integrated distance of the unknown fields $\nabla T$ and $\mathbf{q}$ to the dataset $\mathcal{D}$ can be defined as follows:
\begin{equation}
    \label{eq:minimisation_stronger}
    J( \nabla T, \mathbf q) = \int\limits_\Omega S_{\textrm g} \left( \mathbf \nabla T - \mathbf g^* \right)^2 d\Omega + \int\limits_\Omega S_{\textrm q} \left( \mathbf q - \mathbf q^* \right)^2 d\Omega, 
\end{equation} 
where the data point $\{T^*, \mathbf{g}^*, \mathbf{q}^*\}_{5D}$ or $\{\mathbf{g}^*, \mathbf{q}^*\}_{4D}$ is found as a minimiser of~\eqref{eq:distance_dataset_dd}. The functional~\eqref{eq:minimisation_stronger} can then be minimised by means of the finite element method.

It is worth noting that we chose the integrated distance functional~\eqref{eq:minimisation_stronger} to contain only heat flux and gradient of temperature variables, and not the temperature itself. This is because the relationship between heat flux and gradient of temperature is dictated by the second law of thermodynamics (heat flows from the regions of higher temperature to regions of lower temperature), while any effect of temperature on the heat flux is secondary, as any flux can occur at any temperature within the material dataset.
Therefore, temperature values $T^*$ in the dataset are used solely to find the closest point in~\eqref{eq:distance_dataset_dd}.

\section{Data-driven finite element formulations}
\label{sec:dd_formulations}
The following section explores two data-driven finite element formulations for solving heat transfer problems: the original one (stronger) and the one proposed in this paper (weaker). Both start from the same set of equations: the conservation law \eqref{eq:conservation_energy}, boundary conditions \eqref{eq:bcs} and the integrated distance functional \eqref{eq:minimisation_stronger}. 

\begin{figure} [t]
    \centering
    \includegraphics[width=\linewidth]{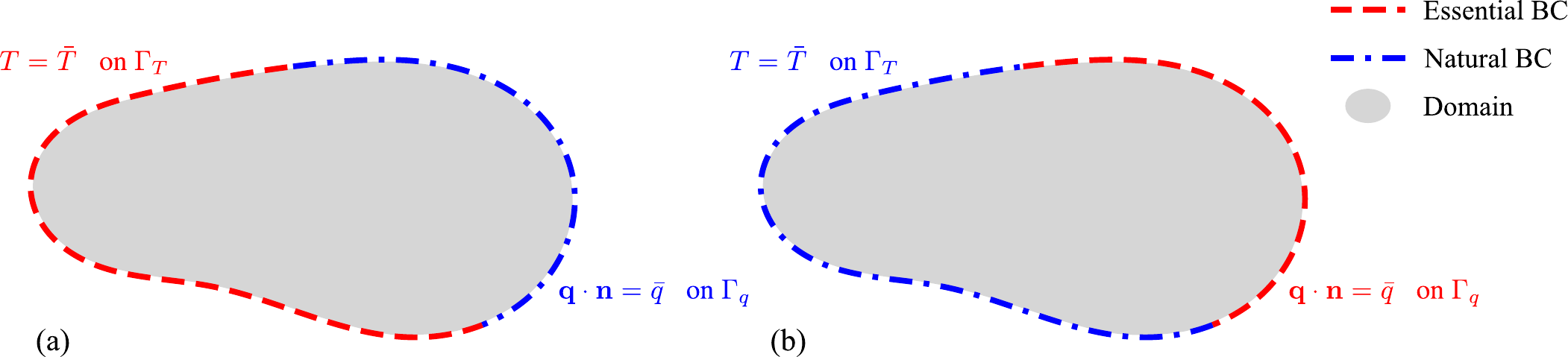}
    \caption{Essential and natural boundary conditions for (a) stronger DD and (b) weaker DD formulations of the heat diffusion problem, cf. Fig.~\ref{fig:potato_bc_general}. }
    \label{fig:potato_discretised_combined}
\end{figure}

The key difference between the stronger and weaker formulations lies in the function spaces used for approximation of the unknown functions, see \autoref{tab:spaces_compared}. While the stronger formulation utilises the Sobolev space $H^1(\Omega)$ for the temperature and the Lebesgue space $\mathbf L^2(\Omega)$ for the flux, the weaker formulation uses the Lebesgue space $L^2(\Omega)$ for the temperature and the Sobolev space $H(\text{div};\Omega)$ for the flux. The weaker formulation is more natural for the heat transfer problem as it enforces continuity of the normal flux component across inner boundaries. Moreover, in the weaker formulation, the Dirichlet and Neumann boundary conditions are enforced as natural and essential, respectively, which is the opposite of the stronger data-driven formulation and the standard FEM, see \autoref{fig:potato_discretised_combined}. 

\begin{table}[b]
    \centering
    \renewcommand{\arraystretch}{1.2} 
    \begin{tabular}{ 
      >{\centering\arraybackslash}p{0.22\textwidth} 
      | >{\centering\arraybackslash}p{0.07\textwidth} 
      | >{\centering\arraybackslash}p{0.3\textwidth} 
      | >{\centering\arraybackslash}p{0.3\textwidth} }
              \hline
      \multirow{2}{*}{Field} & \multirow{2}{*}{Rank} & \multicolumn{2}{c}{Function space} \\ \cline{3-4}
      & & Stronger DD & Weaker DD \\
      \hline
        \rowcolor{gray!25} Temperature & Scalar & $T \in H^1(\Omega) \, : \, T = \bar T \textrm{ on } \Gamma_T$ & $T \in L^2(\Omega)$ \\
        Temperature gradient & Vector & Not used & $\mathbf g \in \mathbf L^2(\Omega)$ \\
        \rowcolor{gray!25} Heat flux & Vector & $\mathbf q \in \mathbf L^2 (\Omega)$ & $\mathbf q \in H(\text{div};\Omega) \,:\, \mathbf{q}\cdot\mathbf{n} = \bar{q} \;\text{on}\; \Gamma_q$ \\
        Lagrange multiplier $\lambda$ & Scalar & $\lambda \in H^1(\Omega) \, : \, \lambda = 0 \textrm{ on } \Gamma_T$ & $\lambda \in L^2(\Omega)$ \\
        \rowcolor{gray!25} Lagrange multiplier $\boldsymbol{\tau}$ & Vector & Not used & $\boldsymbol{\tau} \in  H(\text{div};\Omega) \,:\, \boldsymbol{\tau}\cdot\mathbf{n} = 0 \;\text{on}\; \Gamma_q$ \\
    \end{tabular}
    \caption{Function spaces for the stronger and weaker data-driven formulations. $L^2(\Omega)$ and $\mathbf L^2(\Omega)$ are the Lebesgue spaces for square-integrable scalar and vector functions, respectively. $H^1(\Omega)$ is the Sobolev space for square-integrable scalar functions with square-integrable derivatives. $H(\text{div};\Omega)$ is the Sobolev space for square-integrable vector functions with square-integrable divergence.}
    \label{tab:spaces_compared}
\end{table}

\subsection{Stronger data-driven formulation} \label{sec:stronger_mixed}

Stronger mixed formulation for data-driven approach has been derived, tested and extended for elasticity and inelasticity in~\citep{kirchdoerfer2016data, kirchdoerfer2017data}, while a similar formulation for diffusion problems was discussed in~\citep{nguyen2020variational, kulikova_shvarts_kaczmarczyk_pearce_2021}. Following these works, a DD version for heat transfer including temperature $T$ as one of the dataset $\mathcal{D}$ variables is derived here. 

The solution of the problem~\eqref{eq:conservation_energy}-\eqref{eq:bcs}, supplemented by a dataset generated and handled as discussed in Section~\ref{sec:seepage_intro}, can be found upon the minimisation of the functional \eqref{eq:minimisation_stronger} with the conservation law \eqref{eq:conservation_energy} imposed as a constraint. Therefore, we formulate a Lagrangian:
\begin{equation}
    \label{eq:lagrangian_stronger}
    \mathcal {L} \left(\nabla T, \mathbf q, \lambda \right) = J\left(\nabla T, \mathbf q\right) + \int\limits_\Omega \lambda \left(\nabla \cdot \mathbf q - \gls{source_term}\right) \, d\Omega ,
\end{equation}
where $\lambda$ is the field of scalar Lagrange multipliers. While the Dirichlet boundary condition on temperature (\ref{eq:boundary_p}) is satisfied \textit{a priori}, the Neumann boundary condition on flux (\ref{eq:boundary_q}) can be enforced in the integral sense using the integration by parts of the term representing the constraint in the Lagrangian (\ref{eq:lagrangian_stronger}) and imposing $\lambda = 0$ on $\Gamma_T$, see~\citep{nguyen2020variational}:
\begin{equation}
    \label{eq:integ_by_parts_stronger}
    \begin{aligned}
    \displaystyle
        \mathcal {L}\left(\nabla T, \mathbf q, \lambda\right) &=  J\left(\nabla T,\mathbf q\right) + \int\limits_\Omega {\nabla\lambda} \cdot \mathbf q \, d\Omega - \int\limits_{\Gamma_q} \lambda \mathbf q \cdot \mathbf n \, d\Gamma \\
        &= \int\limits_\Omega S_{\textrm g} \left({\mathbf \nabla T} -\mathbf g^*\right)^2 d\Omega + \int\limits_\Omega S_{\textrm q} \left(\mathbf q - \mathbf q^*\right)^2 d\Omega - \int\limits_\Omega {\nabla\lambda} \cdot \mathbf q \, d\Omega + \int\limits_{\Gamma_q} \lambda \bar{q} \, d\Gamma - \int\limits_\Omega \lambda \gls{source_term} \, d\Omega.
    \end{aligned}
\end{equation}
The solution is found as a stationary point of the Lagrangian at which its variation vanishes:
\begin{equation}
    \label{eq:variational_dd}
    \begin{cases}
        \displaystyle \int\limits_\Omega S_{\textrm g}\left(\nabla T - \mathbf{ g^*}\right) \cdot \nabla \delta T \, d\Omega = 0  & \;\;\;\; \forall \delta T \in H_0^1 (\Omega) \\
        \displaystyle\int\limits_\Omega S_{\textrm q}\left(\mathbf q - \mathbf{q^*}\right) \cdot \mathbf{\delta q} \,~d\Omega + \int\limits_\Omega \mathbf{\nabla \lambda} \cdot \mathbf{\delta q} \, d\Omega = 0  & \;\;\;\; \forall \delta \mathbf q \in \mathbf L^2 (\Omega) \\
        \displaystyle\int\limits_\Omega \mathbf q \cdot \nabla \delta \lambda \,~d\Omega - \int\limits_{\Gamma_q} \bar q \delta \lambda \,~d\Gamma + \int\limits_{\Omega} \gls{source_term} \delta \lambda \, d\Omega = 0  & \;\;\;\; \forall \delta \lambda \in H^1_0 (\Omega),
    \end{cases}
\end{equation}
where $H_0^1(\Omega)$ is a space for square-integrable scalar functions with square integrable gradient and zero value on the boundary $\Gamma_T$, i.e. 
$H_0^1(\Omega) = \{ u \in H^1(\Omega) \, | \, u = 0 \textrm{ on } \Gamma_T \}$.
The unknown trial functions $T, \mathbf{q}, \lambda$ belong to the spaces shown in the ``Stronger DD'' column in \autoref{tab:spaces_compared}. 

The variational problem~\eqref{eq:variational_dd} can be discretised and solved using linear FEM. Here, we used hierarchical shape functions~\citep{ainsworth2003hierarchic}  available in MoFEM~\citep{mofemJoss2020} for both continuous ($T, \lambda$) and discontinuous ($\mathbf{q}$) field approximations, see~\ref{app:fe_strong} and~\citep{kulikova_data-driven_2025} for more details. To satisfy LBB stability conditions~\citep{boffi2013mixed}, the approximation orders have to be chosen as shown in~\autoref{tab:order_compare}, see the ``Stronger DD'' column.

This stronger data-driven formulation will serve as a basis for comparison with the weaker mixed data-driven formulation introduced in the next subsection.

\begin{table}
    \centering
    \renewcommand{\arraystretch}{1.2} 
    \begin{tabularx}{0.8\textwidth} { 
      >{\centering\arraybackslash}X 
      | >{\centering\arraybackslash}Xc
     >{\centering\arraybackslash}X }
        \hline
      \multirow{2}{*}{Field} & \multicolumn{2}{c}{Approximation order} \\ \cline{2-3}
      & Stronger DD & Weaker DD \\
      \hline
     \rowcolor{gray!25} Temperature $T$ & $p$ & $p$ \\
     Gradient of temperature $\mathbf g$ & Not used & $p+1$ \\
     \rowcolor{gray!25} Heat flux $\mathbf q$ & $p - 1$ & $p+1$ \\
     Lagrange multiplier $\lambda$ & $p$ & $p$ \\
     \rowcolor{gray!25} Lagrange multiplier $\boldsymbol{\tau}$ & Not used & $p+1$ \\
    \end{tabularx}
    \caption{Approximation orders to satisfy inf-sub stability conditions; note that $p$ is the approximation order for the temperature field $T$ in both formulations.}
    \label{tab:order_compare}
\end{table}

\subsection{Weaker data-driven formulation} \label{sec:weaker_mixed}

The stronger data-driven formulation derived in \autoref{sec:stronger_mixed} is already mixed, i.e. contains multiple unknown fields belonging to different function spaces. To improve on the original formulation, more natural spaces for the heat transfer problem can be chosen for the unknown fields in the weaker formulation, the advantages are explained later in this subsection.

First, to obtain a weaker mixed formulation \citep{boffi2013mixed}, a new variable $\mathbf{g}$ is introduced as one of the unknown fields using an additional equation:
\begin{equation}
    \label{eq:gradient_temperature}
    \mathbf{g} = \nabla T.
\end{equation}
The closest point search \eqref{eq:distance_dataset_dd} is now rewritten in terms of the new variable $\mathbf{g}$:
\begin{equation}
        \label{eq:distance_dataset_dd_g}
        \textrm{dist} \left(\{T, \mathbf{g} , \mathbf{q} \}, \mathcal{D}\right) = \min_{\{T^*, \mathbf{g^*}, \mathbf{q^*} \} \in \mathcal{D}} \sqrt{S_{\textrm T}\left( T -T^*\right)^2 + S_{\textrm g}\left(\mathbf{g} -\mathbf g^*\right)^2 + S_{\textrm q}\left(\mathbf q - \mathbf q^*\right)^2}.
\end{equation}
Similarly, the integrated distance functional \eqref{eq:minimisation_stronger} becomes:
\begin{equation}
        \label{eq:minimisation_g}
        J\left(\mathbf g,\mathbf q\right) = \int\limits_\Omega S_{\textrm g} \left(\mathbf g -\mathbf g^*\right)^2 d\Omega + \int\limits_\Omega S_{\textrm q} \left(\mathbf q - \mathbf q^*\right)^2 d\Omega.
\end{equation}

Next, the Lagrangian is formed by minimising the integrated distance functional~\eqref{eq:minimisation_g}, and imposing the conservation law~\eqref{eq:conservation_energy} and the new variable definition~\eqref{eq:gradient_temperature} as constraints: 
\begin{equation}
    \label{eq:mixed_lagrangian}
    \mathcal L\left(\nabla T, \mathbf g,\mathbf q, \boldsymbol \tau, \lambda\right) = J\left(\mathbf g,\mathbf q\right) + \int\limits_\Omega \boldsymbol \tau \cdot \left(\mathbf g - \nabla T\right) d\Omega + \int\limits_\Omega \lambda \left(\nabla \cdot \mathbf q - \gls{source_term}\right) d\Omega,
\end{equation}
where $\lambda$ and $\boldsymbol \tau$ are scalar and vector fields of Lagrange multipliers, respectively.

The regularity requirement on the temperature field is weakened by integrating the component containing gradient of temperature $\nabla T$ in~\eqref{eq:mixed_lagrangian} by parts, at the same time satisfying the Dirichlet boundary condition \eqref{eq:boundary_p} as natural and imposing $\boldsymbol \tau \cdot \mathbf n = 0$ on $\Gamma_q$, cf.~\eqref{eq:integ_by_parts_stronger}:
\begin{equation}
\begin{aligned}
    \label{eq:mixed_lagrangian_derivation}
    \mathcal L\left(T, \mathbf g,\mathbf q, \boldsymbol \tau, \lambda\right) =&  \int\limits_\Omega S_{\textrm g} \left( {\mathbf g} -\mathbf g^* \right)^2 d\Omega + \int\limits_\Omega S_{\textrm q} \left( \mathbf q - \mathbf q^* \right)^2 d\Omega \\
& + \int\limits_\Omega \boldsymbol \tau \cdot \mathbf g~d\Omega - \int_{\Gamma_T} \left( \boldsymbol \tau \cdot \mathbf n \right) ~ \bar T ~d \Gamma_T + \int\limits_\Omega \left( \nabla \cdot \boldsymbol \tau \right) ~ T ~ d \Omega \\
& + \int\limits_\Omega \lambda ~ \left( \nabla \cdot \mathbf q \right) ~ d\Omega - \int\limits_\Omega \lambda ~ \gls{source_term} ~ d \Omega.
\end{aligned}
\end{equation}

The choice of function spaces for the unknown fields in the weaker DD formulation, see \autoref{tab:spaces_compared}, results in a lower regularity of the solution, as spaces for $T$ and $\mathbf{g}$ do not require the result to be smooth between the inner boundaries. However, the continuity of the normal flux $\mathbf q$ component is enforced across any inner boundaries. This choice of spaces is more natural for the transport problems such as heat transfer. 

Again, the solution is found as a stationary point of the Lagrangian~\eqref{eq:mixed_lagrangian_derivation} at which its variation vanishes:
\begin{equation}
    \label{eq:mixed_variational}
    \begin{cases}
        \displaystyle \int\limits_\Omega \delta T \left(\nabla \cdot \boldsymbol \tau\right)~d\Omega = 0 &\;\;\; \forall\, \delta T \in L^2(\Omega) \displaystyle \\
        \displaystyle S_{\textrm g} \int\limits_\Omega \delta \mathbf g \cdot \left(\mathbf g - \mathbf g^*\right)~d\Omega + \int\limits_\Omega \delta \mathbf g \cdot \boldsymbol \tau~d\Omega = 0 &\;\;\; \forall \delta \mathbf g \in \mathbf L^2(\Omega) \displaystyle \\
        \displaystyle S_{\textrm q} \int\limits_\Omega \delta \mathbf q \cdot \left(\mathbf q - \mathbf q^*\right)~d\Omega + \int\limits_\Omega \left(\nabla \cdot \delta \mathbf q \right) ~\lambda~d\Omega = 0 &\;\;\; \forall \delta \mathbf q \in  \mathcal{Q}_0 \displaystyle\\
        \displaystyle\int\limits_\Omega \delta \lambda ~ \left(\nabla \cdot \mathbf q\right) ~d\Omega - \int\limits_\Omega \delta \lambda~ \gls{source_term} ~d\Omega = 0 &\;\;\; \forall \delta \lambda \in L^2(\Omega) \displaystyle \\
        \displaystyle\int\limits_\Omega \delta \boldsymbol \tau \cdot \mathbf g + \left(\nabla \cdot \delta \boldsymbol \tau\right)~ T~d\Omega - \int_{\Gamma_T} \left(\delta \boldsymbol \tau \cdot \mathbf n\right)~ \bar T~ d \Gamma = 0 &\;\;\; \forall \delta \boldsymbol{\tau} \in \mathcal{Q}_0,
    \end{cases}
\end{equation}
where $\mathcal{Q}_0$ is a space of square-integrable vectorial functions with square integrable divergence and zero normal component on the boundary $\Gamma_q$, i.e. $\mathcal{Q}_0 = \{\mathbf{v}\in H(\text{div};\Omega) \,|\, \mathbf{v} \cdot \mathbf n = 0 \;\text{on}\; \Gamma_q\}$. The unknown trial functions $T, \mathbf{g}, \mathbf{q}, \lambda, \boldsymbol{\tau}$ belong to the spaces shown in the ``Weaker DD'' column in \autoref{tab:spaces_compared}. 

After discretisation of the domain, each field in the variational problem~\eqref{eq:mixed_variational} can be approximated using hierarchical shape functions, similarly to the stronger case considered above. In particular, we use discontinuous shape functions for $T, \mathbf{g},\, \text{and}\, \lambda$ fields and generalisation of the Brezzi-Douglas-Marini finite element~\citep{boffi2013mixed} for fields $\mathbf{q}\,\text{and}\, \boldsymbol{\tau}$. It is important to note that the approximation orders are assigned to unknown fields as shown in \autoref{tab:order_compare} to satisfy LBB stability conditions. Upon the FE approximation, the saddle point problem~\eqref{eq:mixed_variational} can be represented by a system of linear algebraic equations, see~\ref{app:fe_weak} and~\citep{kulikova_data-driven_2025} for more details.


\subsection{Iterative process and stopping criterion}
\label{sec:iterative_process}
Solving the problem using the data-driven method in either strong or weak formulation is an iterative procedure, where each iteration includes two steps:
\begin{enumerate}
    \item Searching the dataset for the closest material point at every integration point using~\eqref{eq:distance_dataset_dd} in case of the stronger formulation or~\eqref{eq:distance_dataset_dd_g} in case of the weaker formulation.
    \item Solving the finite element problem defined by~\eqref{eq:variational_dd} (stronger formulation) or~\eqref{eq:mixed_variational} (weaker formulation).
\end{enumerate} 

In both cases, the iterative process requires a stopping criterion. The simplest one would be stop the process if the data search with~\eqref{eq:distance_dataset_dd} or \eqref{eq:distance_dataset_dd_g} finds the same material datapoints at every integration point as in the previous iteration. However, in case of a large amount of datapoints being concentrated in a part of the dataset, more iterations do not necessarily provide better results. 

Another option for a stopping criterion is comparing RMS measure of the distance between the field values and the dataset between the two iterations with a prior set tolerance:
\begin{equation}
    \label{eq:rms_D}
    \varepsilon_d\left(T, \mathbf g, \mathbf q\right) = \sqrt{\frac{1}{\mu(\Omega)} \int\limits_\Omega \left(\textrm{dist} \left(\left\{T, \mathbf{g} , \mathbf{q} \right\}, \mathcal{D}\right) \right)^2 \; d\Omega} ,
\end{equation}
where $\mu(\Omega)$ is the area of the domain $\Omega$ in a 2D case. A stopping criterion based on choosing the same closest datapoints at all integration points twice~\citep{kirchdoerfer2016data, nguyen2020variational}, corresponds to setting the tolerance for $\varepsilon_d(T, \mathbf g, \mathbf q)$ to zero in~\eqref{eq:rms_D}.

\subsection{Computational complexity} \label{sec:complexity}

To finish the discussion of the data-driven framework, we  briefly compare the computational complexity of the proposed method with the standard finite element approach to nonlinear problems using the Newton-Raphson scheme~\citep{bonet1997nonlinear}.

For the nonlinear FE formulation, at each Newton-Raphson iteration, a global system is assembled and solved. Assuming an efficient sparse direct solver~\citep{amestoy2000mumps}, the matrix inversion scales as $\mathcal{O}(n_{\text{dof}}^{3/2})$, where $n_{\text{dof}}$ is the number of degrees of freedom. With $N_{\text{NR}}$ Newton-Raphson iterations, the total complexity becomes:
\begin{equation}
    \mathcal{O}(N_{\text{NR}} \cdot n_{\text{dof}}^{3/2}).
\end{equation}

In contrast, the data-driven formulations have a constant diffusivity matrix requiring only a single matrix inversion: $\mathcal{O}(n_{\text{dof}}^{3/2})$. 
However, an iterative process which includes searching for a closest material datapoint for each integration point is performed.
In our implementation, a nearest-neighbour search is carried out at every integration point using an R-tree \citep{guttman1984r}, yielding a search cost of $\mathcal{O}(n_{\text{int}} \cdot \log n_{\text{mat}})$,
where $n_{\text{int}}$ is the number of integration points and $n_{\text{mat}}$ the number of material data points. With $N_{\text{DD}}$ iterations, the total complexity is:
\begin{equation}
    \label{eq:complexity_dd}
    \mathcal{O}(n_{\text{dof}}^{3/2}) + \mathcal{O}(N_{\text{DD}} \cdot n_{\text{int}} \cdot \log n_{\text{mat}}).
\end{equation}

The resulting computational cost difference between the standard and data-driven formulations depends on the size of the problem ($n_{\text{dof}}$ and $n_{\text{int}}$), the size of the material dataset ($n_{\text{mat}}$), and the number of iterations required for convergence ($N_{\text{NR}}$ or $N_{\text{DD}}$). 
Larger problems with more degrees of freedom and integration points might be solved faster with the data-driven approach, as it avoids the need for multiple matrix inversions, see~\citep{eggersmann2021efficient}, especially if the material dataset contains a small number of points. On the contrary, smaller problems with fewer degrees of freedom and integration points will likely perform faster with the standard approach, especially if the material dataset is large.

Therefore, the bottleneck of the data-driven approach is the search through the material dataset, which can be minimised by using efficient data structures (e.g., R-trees) and search algorithms.
Additionally, to ensure the minimum number of integration points for a given accuracy, the error indicators leading to adaptive refinement strategies, which will be discussed in \autoref{sec:error_estimates}, are advantageous. 

\section{Verification and comparison of stronger and weaker DD formulations}
\label{sec:verification}

\begin{figure}[t]
    \centering
    \includegraphics[width=\linewidth]{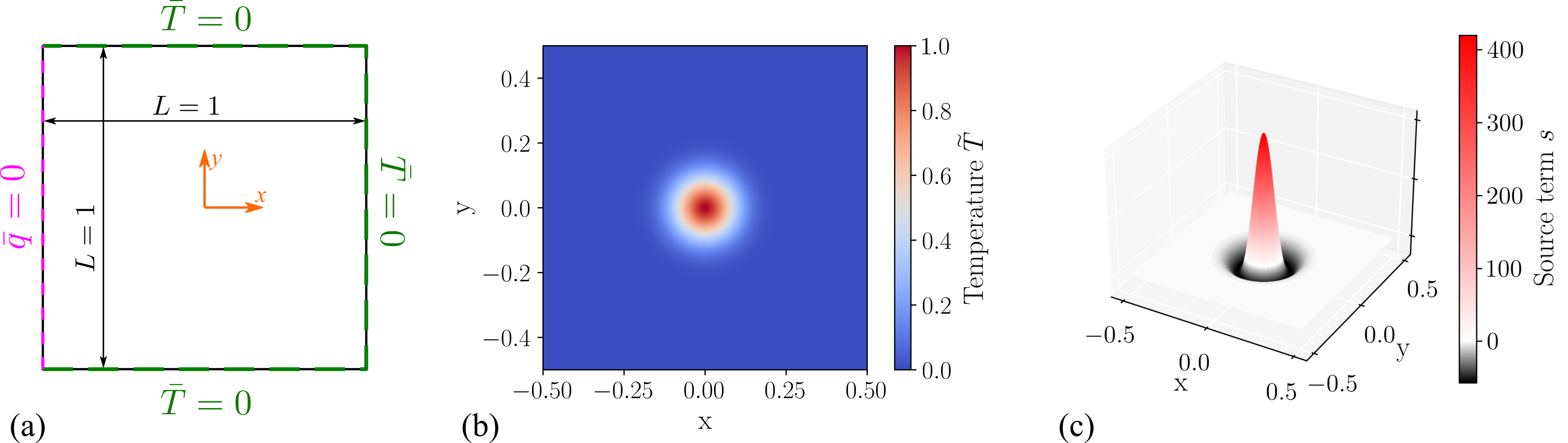}
    \caption{Example Exp-Hat: analysis setup which corresponds to the exact solution $ \tilde T = e^{-100(x^2 + y^2)}  \cos \pi x \cos \pi y$. (a) boundary conditions; (b) temperature solution $T$; (c) source term $s$.}
    \label{fig:numerical_example_2_mexi_hat_general_setup}
\end{figure}

To verify the implementation of the data-driven formulations discussed in section~\ref{sec:dd_formulations}, a diffusion problem in a square 2D domain is considered, see \autoref{fig:numerical_example_2_mexi_hat_general_setup}. The same example will be used in Section~\ref{sec:error_estimates} to demonstrate the effectiveness of the adaptive refinement driven by the error estimators discussed in the same section.

For testing purposes, we constructed the problem for a given field of temperature $\tilde{T}(x,y)$ in a square domain, see \autoref{fig:numerical_example_2_mexi_hat_general_setup}: 
\begin{equation}
    \label{eq:analytical_solution_hat}
\tilde T(x,y) = e^{-100 \left(x^2 + y^2\right)}  \cos \pi x \cos \pi y,
\end{equation}
from which the applied source term $\gls{source_term}(x, y)$ and boundary conditions for heat flux $\bar{q}$ and temperature $\bar{T}$ are derived. 
The example was chosen due to the high localised gradients which cause numerical difficulties and high finite element approximation errors with low order approximations. 

\begin{figure}  [t]
    \centering
    \includegraphics[width=\linewidth]{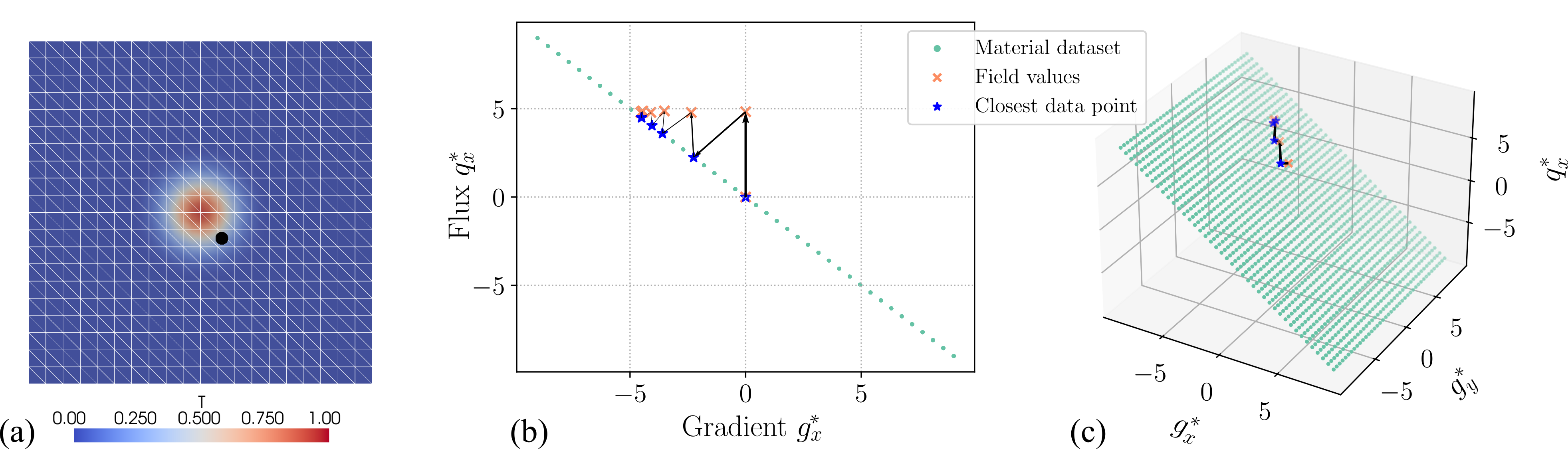}
    \caption{Example Exp-Hat: (a) observed integration point marked black in the domain; (b-c) 2D and 3D paths of the observed integration point, which includes finding the closest point in the dataset (``Closest data point'') and projecting it onto a set where conservation laws and boundary conditions are satisfied (``Field values''). The problem is solved using weaker DD formulation~\eqref{eq:mixed_variational} with mesh size $h=0.05$, approximation order $p=2$, and dataset including $41^2$ material datapoints.}
    \label{fig:integration_journey}
\end{figure}

For this example, the material dataset $\mathcal D_{4D} = \{g_x^*, g_y^*, q_x^*, q_y^*\}$, see \autoref{fig:artificial_datasets} (a), was generated with thermal conductivity $k = 1$, $g_x$ and $g_y$ in range $[-9,9]$, and $41^2$ points, see details of dataset creation described in \ref{app:regular_dataset}. The material dataset $\mathcal D_{4D}$ in this example does not contain values for $T^*$, and therefore the terms containing $T^*$ in~\eqref{eq:distance_dataset_dd} or~\eqref{eq:distance_dataset_dd_g} are ignored. Using this example dataset, the iterative process discussed in subsection~\ref{sec:iterative_process} is shown for one integration point in~\autoref{fig:integration_journey}.  In the above all units are consistent.


\begin{figure} [t]
    \centering
    \includegraphics[width=\linewidth]{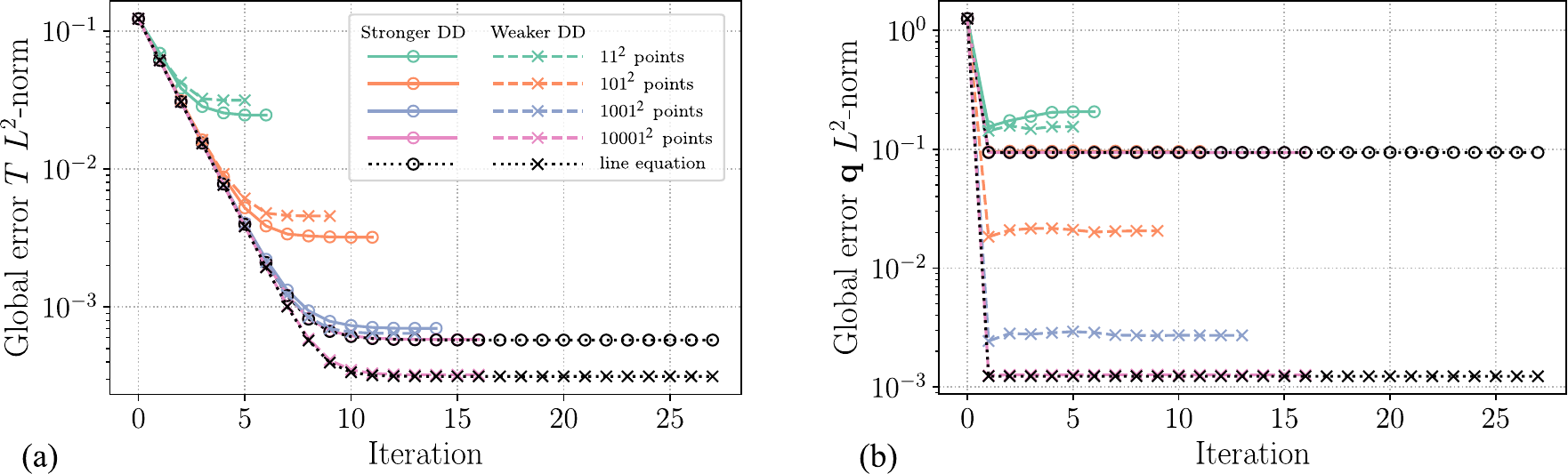}
    \caption{Example Exp-Hat: Convergence of the global (a) temperature and (b) flux error $L^2$-norms w.r.t. the number of points in a material dataset $\mathcal D_{4D}$ ($h=0.05$, $p=2$). Line equation is analytical model of dataset with infinite number of datapoints and no noise, i.e. ``saturated'' dataset, see \ref{app:closest_point}.} 
	\label{fig:c5_errors_with_iterations}
\end{figure}

Next, the material dataset $\mathcal D_{4D}$ is generated multiple times with increased density of datapoints. The global error $L^2$-norm of the $T$ and $\mathbf{q}$ fields obtained in the data-driven formulations is computed with respect to the exact solution. Furthermore, the results are compared to the case of the ``saturated'' dataset, that is, when the data search \eqref{eq:distance_dataset_dd} or \eqref{eq:distance_dataset_dd_g} is replaced by finding a closest point on a line $\mathbf q^* = - \mathbf g^*$, see \ref{app:closest_point} for more details. As noise is not present in the material dataset, the resulting global error converges with the increase in the number of material datapoints in the dataset $\mathcal D_{4D}$, see \autoref{fig:c5_errors_with_iterations}. 
Note that the stronger and weaker DD formulations converge to different values of the global error norms. In particular, the resulting error in the case of the ``saturated'' dataset  corresponds to the finite element approximation error, which differs between the stronger and weaker formulations.

\begin{table} [b!]
    \centering
    \renewcommand{\arraystretch}{1.2} 
    \begin{tabularx}{0.8\textwidth} { 
      >{\centering\arraybackslash}X 
      | >{\centering\arraybackslash}X
       >{\centering\arraybackslash}X }
      \multirow{2}{*}{Error norm} & \multicolumn{2}{c}{Convergence order} \\ \cline{2-3}
      & Standard FEM & Mixed FEM \\
      \hline
     $|| \tilde T - T^h||_{L^2(\Omega)} $ & $p+1$ & $p+1$ \\
      \hline
     $|| \tilde{\mathbf q} - \mathbf q^h||_{\mathbf{L}^2(\Omega)}$ & $p$ & $p+2$ \\
    \end{tabularx}
    \caption{\textit{A priori} error estimates for standard and mixed FEM formulations without the data-driven approach~\citep{boffi2013mixed}. Note that \gls{order_T} is the approximation order of the temperature field in both formulations.}
    \label{tab:a_priori_compare}
\end{table}

\begin{figure}[b!]
    \centering
    \includegraphics[width=\linewidth]{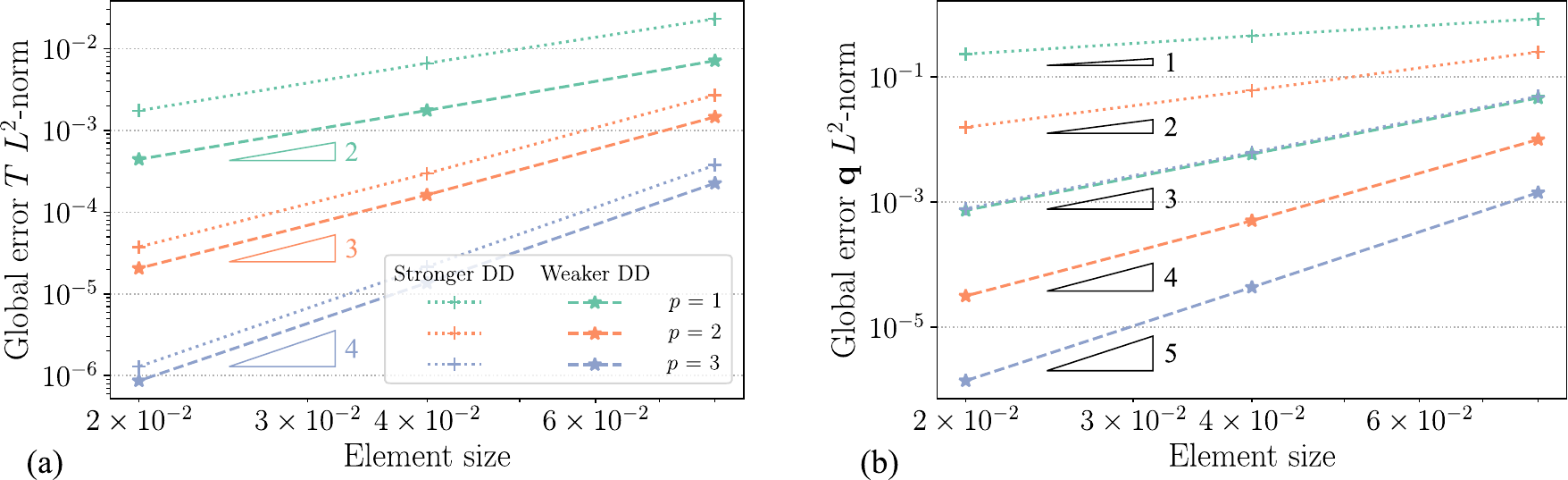}  
    \caption{Example Exp-Hat: Comparison of the convergence w.r.t. element size $h$ and approximation order $p$ of weaker and stronger DD solutions  with ``saturated'' dataset to the exact solution. (a) global temperature $T$ error $L^2$-norm; (b) global flux $\mathbf q$ error $L^2$-norm. } 
	\label{fig:c5_mixed_hat_all_elem_size_compare}
\end{figure}

The convergence of the DD solutions using the ``saturated'' dataset to the exact solution with decreasing element size $h$ and increasing approximation order $p$ is shown in \autoref{fig:c5_mixed_hat_all_elem_size_compare}.
The agreement between the convergence orders apparent in the slopes in \autoref{fig:c5_mixed_hat_all_elem_size_compare} and {\it a priori} error estimates shown in \autoref{tab:a_priori_compare} demonstrate that the stronger and weaker data-driven approaches with ``saturated'' datasets have the same convergence rates as the standard and mixed FE formulations without the data-driven approach~\citep{boffi2013mixed}.

The temperature $T$ field converges to the exact solution with the same order for both stronger and weaker formulations, however, since $T \in L^2$ in the weaker formulation, approximation order $p=0$ is possible, i.e.  temperature is allowed to be constant per element, see \autoref{tab:comparison_fields}. Note that while the approximation of the field $T$ in $L^2$ space allows for jumps between the element boundaries, these jumps decrease with increasing approximation order, provided that the source term applied to the problem is sufficiently smooth. On the other hand, the convergence of the heat flux $\mathbf q$ in the weaker formulation is two orders higher compared to the stronger formulation, see \autoref{tab:comparison_fields}. In particular, the resulting profile of the flux with $p=0$ in the weaker formulation is very similar to the result with $p=2$ in the stronger formulation.

\begin{table}[t!]
    \centering
    \renewcommand{\arraystretch}{1.2} 
    \begin{tabular}
        { 
            >{\centering\arraybackslash}m{0.08\textwidth} |
      >{\centering\arraybackslash}m{0.2\textwidth} 
    >{\centering\arraybackslash}m{0.2\textwidth} |
    >{\centering\arraybackslash}m{0.2\textwidth} 
    >{\centering\arraybackslash}m{0.2\textwidth} }
        \multirow{2}{*}{order} & \multicolumn{2}{c|}{Stronger DD}  & \multicolumn{2}{c}{Weaker DD} \\
        \cline{2-5}
        & Temperature ($H^1$) & Flux ($\mathbf L^2$) magnitude & Temperature ($L^2$) & Flux ($H(\textrm{div})$) magnitude \\   
        \hline  
        $p=0$ & & & 
        \includegraphics[width=0.99\linewidth]{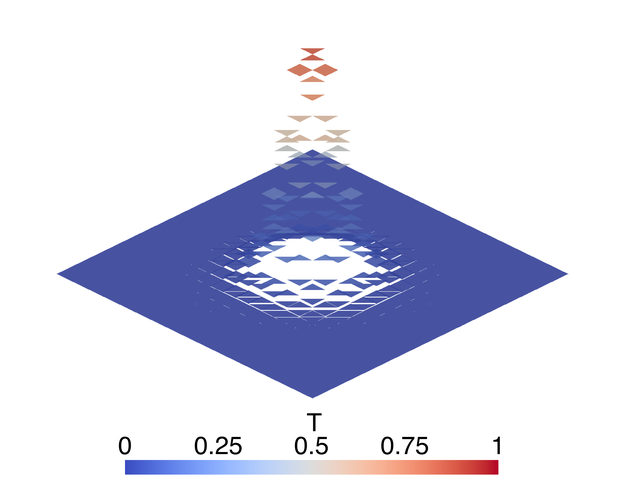} & 
        \includegraphics[width=0.99\linewidth]{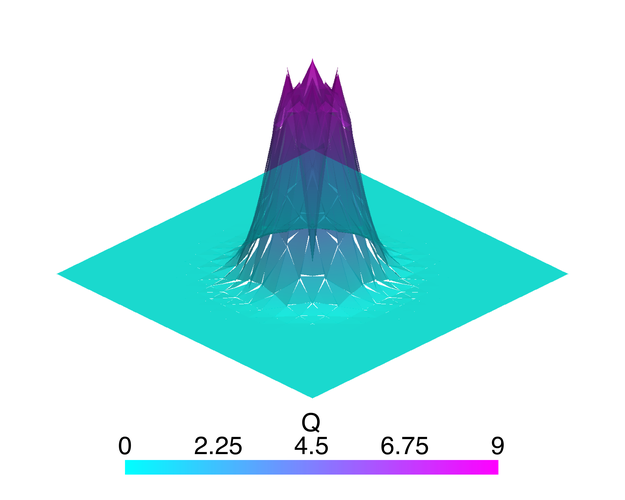} \\
        $p=1$ &
        \includegraphics[width=0.99\linewidth]{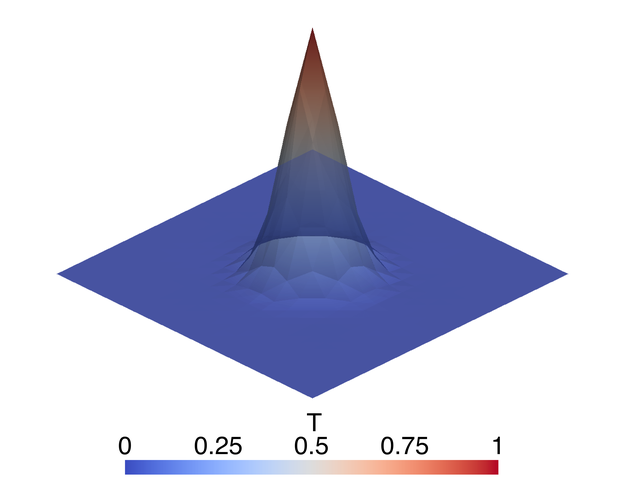} & 
        \includegraphics[width=0.99\linewidth]{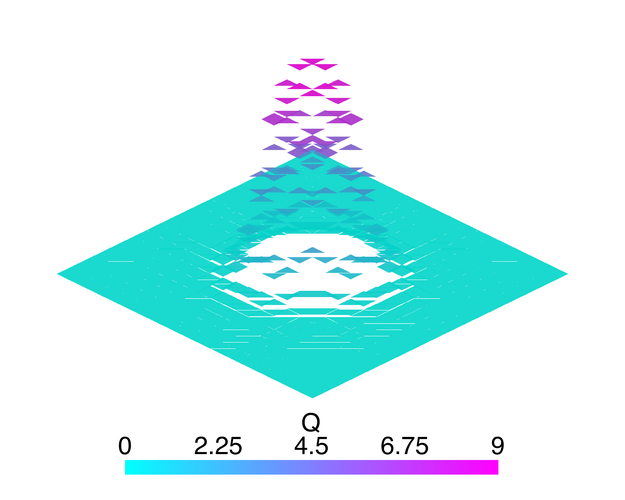} & 
        \includegraphics[width=0.99\linewidth]{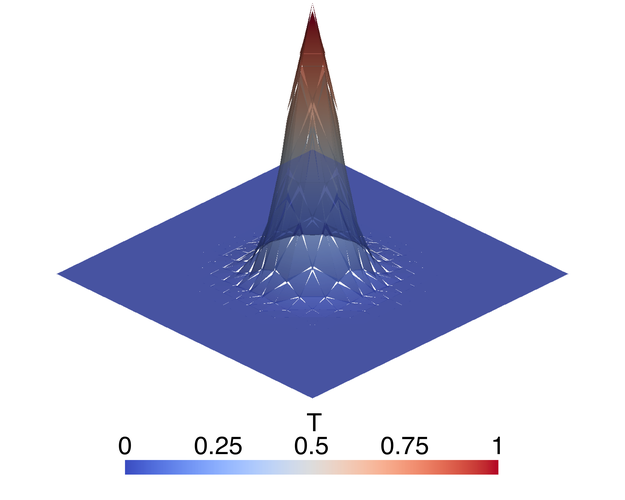} & 
        \includegraphics[width=0.99\linewidth]{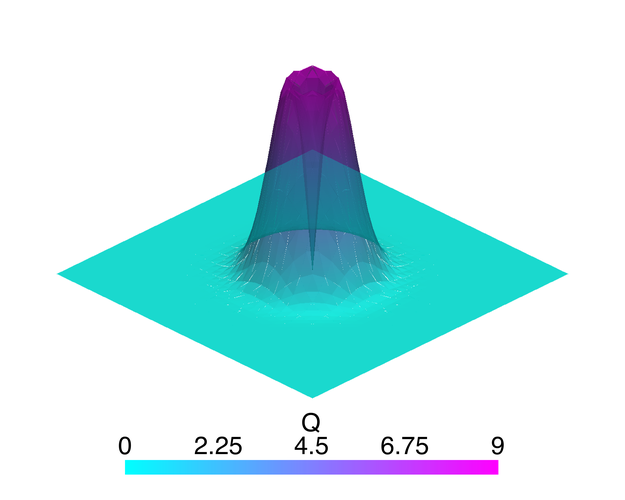} \\
        $p=2$ &
        \includegraphics[width=0.99\linewidth]{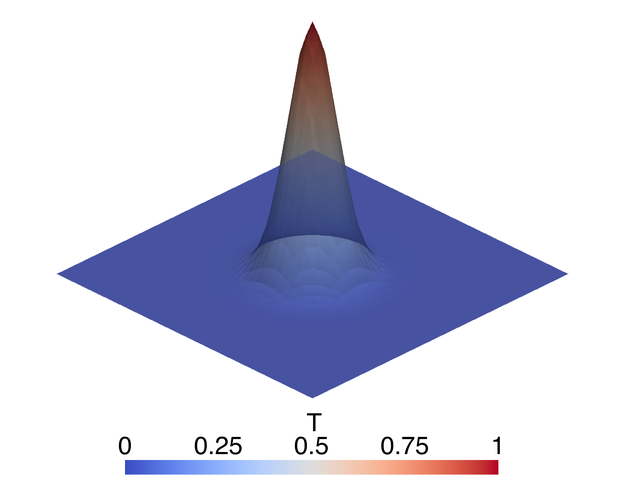} & 
        \includegraphics[width=0.99\linewidth]{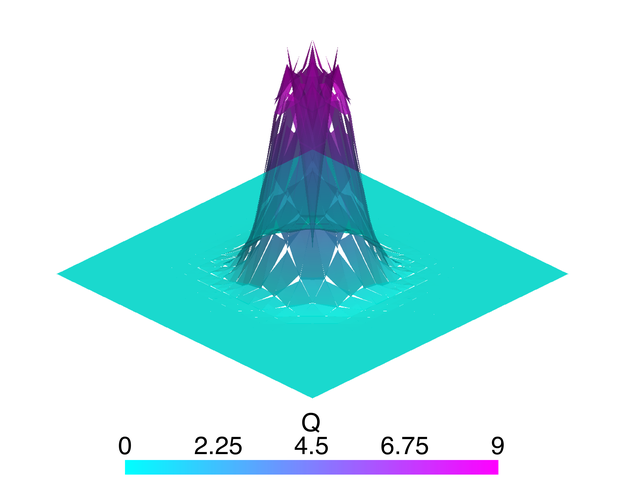} & 
        \includegraphics[width=0.99\linewidth]{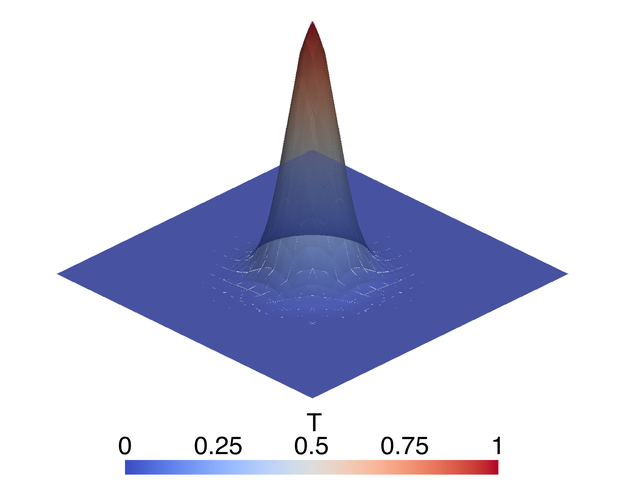} & 
        \includegraphics[width=0.99\linewidth]{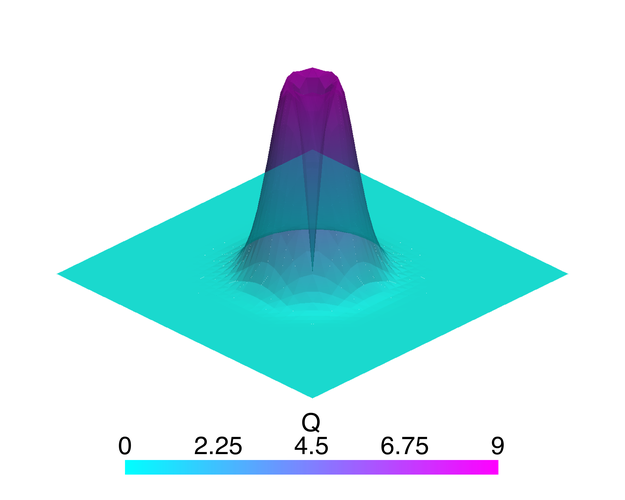} \\
    \end{tabular}
    \caption{Comparison of fields for different approximation orders between the stronger and weaker DD formulations ($h=0.05$). Note that \gls{order_T} is the approximation order of the temperature field in both formulations.}
    \label{tab:comparison_fields}
\end{table}


\section{Adaptive $hp$-refinement with \textit{a posteriori} error indicators} \label{sec:error_estimates}

Numerical solution obtained with the finite element method contains FE approximation errors. Refining mesh or increasing approximation order on the elements with high values of error indicators aims to reduce the FE approximation errors with minimal increase in the problem size and simulation runtime. Adaptive refinement algorithm will be discussed in the following subsections, considering numerical errors and uncertainties related to the material dataset, addressed in \autoref{sec:dd_indicators}.

\subsection{Finite element approximation error indicators}

Multiple ways to calculate FE approximation error indicators and estimators were proposed in literature \citep{oden1989toward, repin2008posteriori, gratsch2005posteriori}. 
For the weaker mixed DD formulation considered here, due to the temperature $T$ and flux $\mathbf q$ belonging to different function spaces, error indicators and estimators with little extra computational cost are available \citep{ainsworth2008posteriori, braess1996posteriori, carstensen1997posteriori}.
The particular {\it a~posteriori} error indicators and error estimator calculated in this paper follow \cite{braess1996posteriori}. 

Error indicators ``indicate'' where the approximation errors discoverable by that particular indicator are the highest, however, any one error indicator might not capture all of the finite element approximation errors on its own. Therefore, the error estimator introduced in this section is composed of error indicators associated with temperature gradient, flux divergence, and jumps of values across the inner boundaries between elements.

\subsubsection{Finite element error indicators} \label{sec:error_indicators}

The first error indicator is associated with the temperature gradient $\nabla T$. Since temperature $T$ is approximated in $L^2$ space, the field is not enforced to be continuous. Nevertheless, given that in the FE approach the fields are approximated using polynomial shape functions, even if the temperature field is not continuous between elements, the gradient of the corresponding polynomial approximation $T^h$ can be readily computed at each element's integration points {\it a posteriori}. The temperature gradient can then be compared to the approximated field $\mathbf g$, which represents the gradient of temperature as well, see \eqref{eq:gradient_temperature}. The error indicator $\eta_e(\nabla T, \mathbf{g})$ is computed over a finite element $\Omega_e$ as follows:
\begin{equation}
    \label{eq:error_indicator_grad}
    \eta_e \left(\nabla T, \mathbf{g}\right)= ||\nabla T - \mathbf{g}||_{\Omega_e}^2,
\end{equation}
where by $||\cdot||_{\Omega_e}$ we mean the $L^2$-norm computed over an element $\Omega_e$.

\begin{figure}  [tb]
    \centering
    \includegraphics[width=\linewidth]{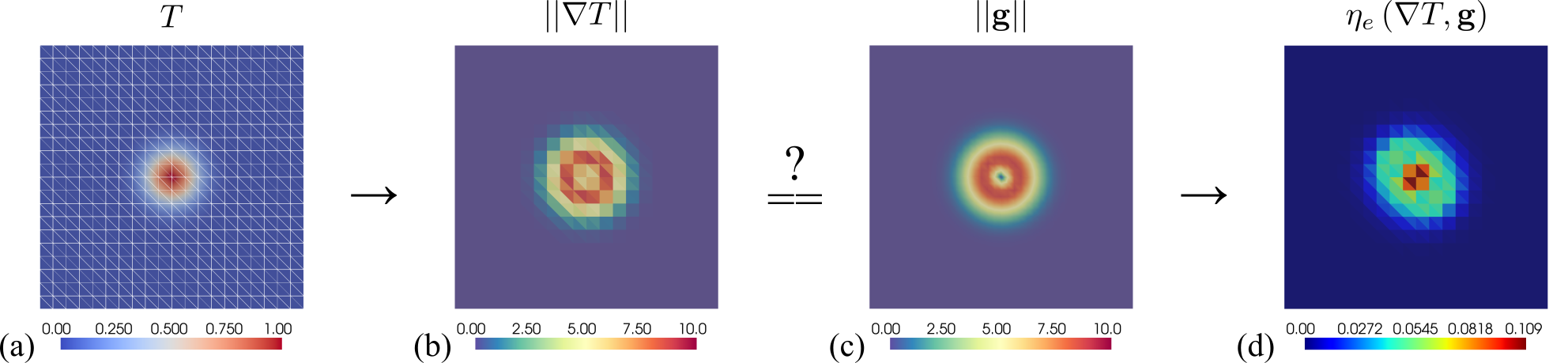}
    \caption{Example Exp-Hat: (a) the temperature field $T$; comparison of the magnitudes of (b) the gradient of temperature $\nabla T$ and (c) the gradient field $\mathbf g$ (approximation order 
    $p = 1$ and element size $\gls{element_size}=0.05$); (d) the error indicator associated with temperature gradient $\eta_e \left(\nabla T, \mathbf{g}\right)$, see \eqref{eq:error_indicator_grad}.}
    \label{fig:c3_grad_T_Q}
\end{figure}

The magnitudes of the fields for the problem considered in section~\ref{sec:verification}, shown in \autoref{fig:c3_grad_T_Q}(b) and (c), in the ideal case, should have the same values, since they both represent the gradient of temperature. However, due to the finite element approximation satisfying the relation \eqref{eq:gradient_temperature} in a weak sense, the values differ and produce the error indicator $\eta_e$, see \autoref{fig:c3_grad_T_Q}(c). Moreover, while the temperature field, see \autoref{fig:c3_grad_T_Q}(a), is approximated with order $p$ shape functions, the gradient field $\mathbf g$ is approximated with the higher approximation order, $p+1$. Therefore, the resulting field $\mathbf{g}$ is smoother than the computed gradient of temperature $\nabla T$.

\begin{figure}  [b!]
    \centering
    \includegraphics[width=\linewidth]{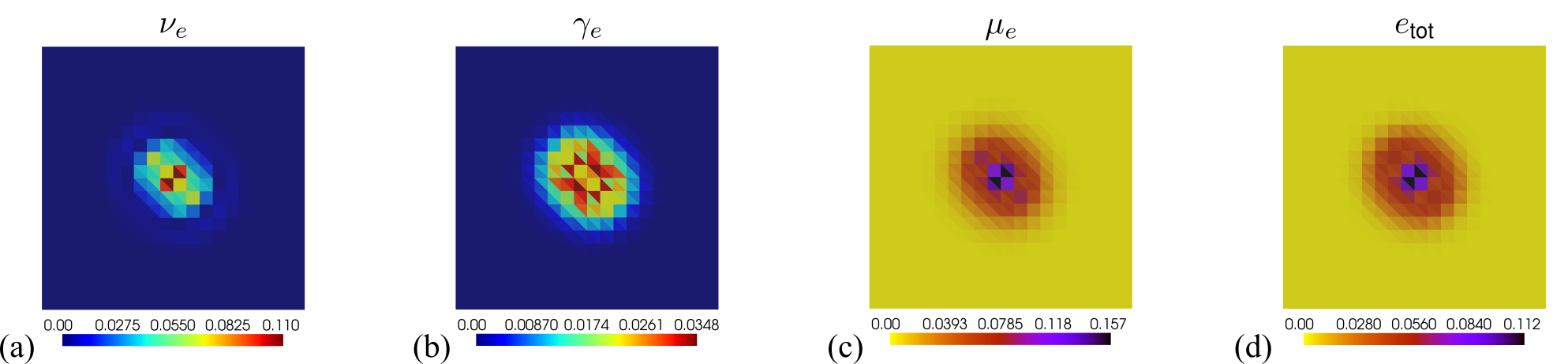}
    \caption{Example Exp-Hat: Visual representation of (a) the divergence error indicator \eqref{eq:error_indicator_div}, (b) the jump error indicator \eqref{eq:error_indicator_jump_sum}, (c) a combined error estimator $\mu_e$ \eqref{eq:error_estimator} and (d) total error w.r.t. the exact solution $e_\text{tot}$ \eqref{eq:total_error} on element basis.}
    \label{fig:c3_err_combined}
\end{figure}

The second error indicator $\nu_e$ measures the difference between the source term \gls{source_term} and the divergence of the flux $\nabla \cdot \mathbf q$, which corresponds to the conservation of energy \eqref{eq:conservation_energy}, and is computed over a finite element $\Omega_e$ as follows:
\begin{equation}
    \label{eq:error_indicator_div}
    \nu_e \left(\mathbf q\right) = h_1 ||\gls{source_term} - \nabla \cdot \mathbf q||_{\Omega_e}^2,
\end{equation}
where $h_1=\sqrt{|\Omega_e|}$ is the representative size of the element $\Omega_e$. The values of the indicator $\nu_e(\mathbf q)$ are shown in \autoref{fig:c3_err_combined}(a).

The final error indicator is associated with temperature jumps across the inner boundaries, see \autoref{tab:comparison_fields} for a visual representation. The jump error indicator $\gamma_e(T)$ for an element is computed as follows:
\begin{equation}
    \label{eq:error_indicator_jump_sum}
    \gamma_e(T) = \sum_{l=1}^{n_{\Gamma_e}} h_2^{-1/2} ||J(T)||_{0,l},
\end{equation}
where $J(T)$ is the jump of the temperature field $T$ across one of internal edges $\Gamma_l$ belonging to an element $\Omega_e$ and $n_{\Gamma_e}$ is the total number of internal edges associated with this element. Here, $h_2 = \frac{\sqrt{|\Omega_{e1}} + \sqrt{|\Omega_{e2}|}}{2}$ is the representative size of the element, while $|\Omega_{e1}|$ and $|\Omega_{e2}|$ are the areas of the elements $\Omega_{e1}$ and $\Omega_{e2}$ sharing the edge $\Gamma_l$. However, it is important to note that the exact solution for the field $T$ does not allow discontinuities. Thus, the error indicator~\eqref{eq:error_indicator_jump_sum}, see \autoref{fig:c3_err_combined}(b), can be used to identify elements where the jump in the temperature field is too high and where the refinement might be needed to achieve convergence towards the exact continuous solution. 

The error estimator~\citep{braess1996posteriori} is a combination of the error indicators associated with temperature gradient~\eqref{eq:error_indicator_grad}, flux divergence~\eqref{eq:error_indicator_div}, and jumps across the inner boundaries~\eqref{eq:error_indicator_jump_sum}:
\begin{equation}
    \label{eq:error_estimator}
    \begin{aligned}
        \mu_e = \left(\eta_e^2 + \nu_e^2 + \gamma_e^2 \right)^{1/2}.
    \end{aligned}
\end{equation}
The error estimator~\eqref{eq:error_estimator} can also be extended to calculate the lower and upper bounds of the numerical error \citep{braess1996posteriori}, which, however, is not in the scope of this paper. \autoref{fig:c3_err_combined}(c) demonstrates the resulting error estimator compared to the total error w.r.t. analytical solution, \autoref{fig:c3_err_combined}(d):
\begin{equation}
    \label{eq:total_error}
       e_\text{tot} = || T - \tilde T ||_{1, \Omega_e} + || \mathbf q - \tilde{\mathbf q} ||_{0, \Omega_e},
\end{equation}
which is the sum of the $H^1$-norm of the temperature error and $L^2$-norm of the flux error. 

\subsubsection{Error indicators with imperfect material dataset}

The error estimator~\eqref{eq:error_estimator} can be used to identify elements where the solution is not well approximated and where the refinement is likely to improve the numerical solution accuracy. However, the data-driven approach uses a dataset which is rarely as perfect as shown in the previous section. Therefore, the error estimator~\eqref{eq:error_estimator} can only be seen as another error indicator for the DD approach since it does not take into account the uncertainties related to the material dataset. Nevertheless, for the purpose of brevity, $\mu_e$ will continue to be referred to as an error estimator in the following sections.

\begin{figure}[tb]
    \centering
    \includegraphics[width=\linewidth]{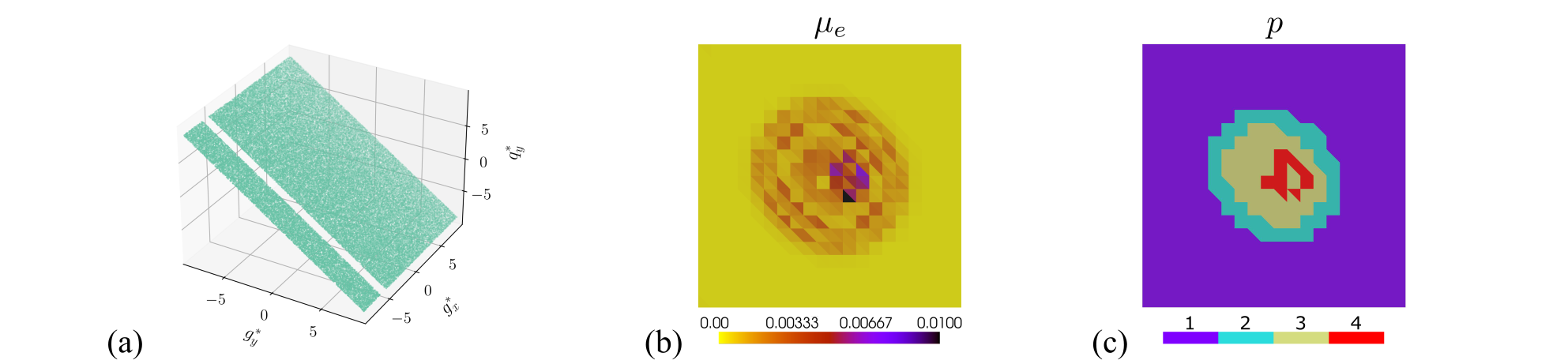}
    \caption{(a) \textit{Tweaked} dataset created with $10^5$ material datapoints, having a set of datapoints with values of the temperature gradient in $x$-direction in the range $[-6,-5]$ removed. (b) Error estimator $\mu_e$ and (c) approximation order $p$ per element for the weaker data-driven formulation with missing data after three iterations of the adaptive order refinement ($t_p = 1.5$), see  \autoref{alg:c5_p_refinement}.} 
    \label{fig:c5_hat_missing_err_est_full_ord_ref}
\end{figure}

To demonstrate the limitations of the error estimator $\mu_e$ in case of an imperfect dataset, a \textit{tweaked} dataset is created with $10^5$ material datapoints 
by independently generating random values in range $[-9, 9]$ for the components of the temperature gradient $g_x^*$ and $g_y^*$ using a uniform distribution, and calculating $q_x^*$ and $q_y^*$ with $k=1$. Additionally, a set of datapoints with values of temperature gradient in $x$-direction in the range $[-6,-5]$ is removed, resulting in a controlled form of missing data, see~\autoref{fig:c5_hat_missing_err_est_full_ord_ref}(a). An adaptive order refinement, see \autoref{alg:c5_p_refinement}, is performed, increasing the approximation order by one on elements where the error estimator is higher than the average error estimator:
\begin{equation}
    \label{eq:average_error_estimator}
    \mu_{\textrm{avg}} = \frac{1}{N} \sum_{e = 1}^N \mu_e,
\end{equation}
where $N$ is the total number of elements, and $\mu_e$ is the value of the error estimator for the element $e$.

\begin{algorithm} 
    \caption{Adaptive $p$-refinement driven by error estimator}
    \label{alg:c5_p_refinement}
    \begin{algorithmic}[1]
        \State Initialize with a uniform approximation order, e.g. $p=1$.
        \Repeat
            \State Solve the problem \eqref{eq:mixed_variational}.
            \State Calculate the error indicators and estimator \eqref{eq:error_estimator}.
            \State Calculate the average error estimator~\eqref{eq:average_error_estimator}.
            \For{each element $e$ in \{1, \dots, N\}}
            \If{$\mu_e > {\mu}_{\textrm{avg}} \gls{err_est_order_tolerance}$}
                \State $p_e := p_e + 1$
            \EndIf
            \EndFor
        \Until{a specified number of refinements is reached.}
    \end{algorithmic}
\end{algorithm}

Conventionally, adaptive $p$-refinement delivers an exponential convergence for problems without singularities \citep{gui1986h}, such as the problem considered in this section. 
However, the error estimator~\eqref{eq:error_estimator} and the approximation order $p$ per element shown in \autoref{fig:c5_hat_missing_err_est_full_ord_ref}(b-c), demonstrate that increasing the approximation order on the elements with high error estimator does not necessarily lead to a more accurate result. In particular, elements with highest error estimator are also those with the highest approximation order, and the same elements have the temperature gradient in $x$-direction in the range $[-6,-5]$, i.e. values that are missing from the material dataset. 

Therefore, the error estimator $\mu_e$ \eqref{eq:error_estimator} lacks the information about whether the errors are related to the FE approximation or to the material dataset if imperfections such as missing data are present. To overcome this, the relation of the errors to the material dataset should also be taken into account in the adaptive refinement process, which is discussed in the following subsection. 

\subsection{Data-driven error indicators} \label{sec:dd_indicators}

When elements with elevated error indicators due to the material dataset issues are refined (both in terms of element size and approximation order), more accurate result is not guaranteed. In particular, increasing the polynomial order of shape functions can lead to unwanted overfitting of the result to an imperfect material data. To prevent this, additional error indicators are proposed and included in the adaptive refinement: the average distance to the material dataset and the standard deviation of the distance to the dataset per element. 


\subsubsection{Distance to the material dataset per element} \label{sec:c5_distance_average}

The distance of the results to the material dataset, which will be denoted hereinafter as $d$, is calculated while finding the closest points in the dataset $\mathcal{D}$ for every integration point using \eqref{eq:distance_dataset_dd} or \eqref{eq:distance_dataset_dd_g}. To identify problematic elements, average of the distances per element is calculated as follows:
\begin{equation}
    \label{eq:c5_distance_average}
    d_{ave}^e = \frac{1}{G^e}\displaystyle \sum_{g = 1}^{G^e} d_g^e,
\end{equation}
where $d_g^e$ is the distance~\eqref{eq:distance_dataset_dd_g} of the results to the material dataset at an integration point $g$ and $G^e$ is the number of integration points per the considered element $e$. For visualisation of the average distance to the material dataset per element, see \autoref{fig:n_missing_hat_dist}(a).

If the average distance to the material dataset is \textit{too high}, the element is deemed to be poorly informed of the material properties and further approximation order refinement would be overfitting the result. More discussion on what is \textit{too high} is given in subsection~\ref{sec:distance_criteria}. 

\subsubsection{Standard deviation of the distance to the material dataset per element} \label{sec:c5_distance_std}

\begin{figure} [tb]
    \centering
    \includegraphics[width=\textwidth]{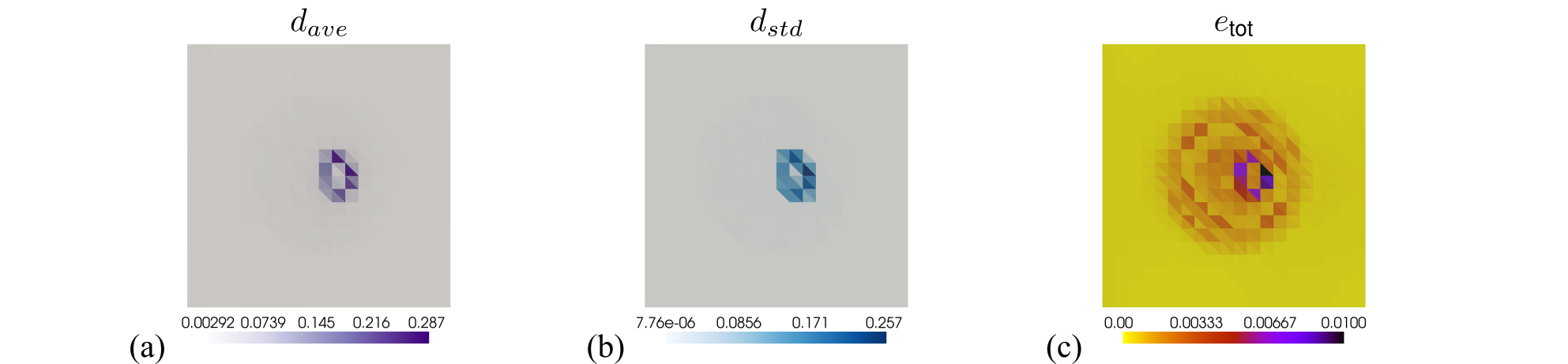}
    \caption{(a) Average distance to the material dataset per element, (b) standard deviation of the distance to the material dataset per element, and (c) total error w.r.t. the exact solution per element for the weaker data-driven formulation using adaptive order refinement algorithm and the \textit{tweaked} dataset with missing data.}
    \label{fig:n_missing_hat_dist}
\end{figure}

On the other hand, let us consider an element one part of which is well informed of the material properties, and another part is not. 
Consider the example in \autoref{fig:c5_hat_missing_err_est_full_ord_ref}, where data is missing in a localised area. 
Due to the size of the elements highlighted by the error estimator $\mu_e$, the missing material datapoints may affect only a part of an element, i.e. some of the integration points may have a large distance to the dataset~\eqref{eq:distance_dataset_dd_g}, while others may have a small distance, while the average distance to the dataset $\gls{d_ave}$ might be similar to other elements. To identify such a case, the \textit{standard deviation} of the distance to the material dataset per element is calculated as:
\begin{equation}
    \label{eq:c5_distance_std}
    \gls{d_std} = \sqrt{\frac{1}{G^e}\displaystyle \sum_{g = 1}^{G^e} \left(d_g^e - \gls{d_ave}\right)^2},
\end{equation}
where $\gls{d_ave}$ is defined by \eqref{eq:c5_distance_average}. For visualisation of the standard deviation of the distance to the material dataset per element, see \autoref{fig:n_missing_hat_dist}(b).

It can be seen in \autoref{fig:n_missing_hat_dist}(c) that after the adaptive order refinement, the total error w.r.t. the exact solution is still present for elements where the dataset is far from the field values. Highlighting the high standard deviation of the distance per element, we can also determine which parts of these elements are better informed (i.e. have smaller distance to the material dataset) of the material properties than others. Therefore, $h$-refining such elements may be beneficial for better understanding the influence of the material dataset on the results.

\subsubsection{Comparative measurement of the distance to the material dataset} \label{sec:distance_criteria}

To address the question of which values of the distance to the material dataset are \textit{too high}, a comparative measure \gls{d_ave_criteria} with the same units as distance in~\eqref{eq:distance_dataset_dd_g} is needed, which would represent RMS of the field values relevant for the material data search. Such measure \gls{d_ave_criteria} can either be calculated locally for each element or globally for the whole domain. Since here we are interested in high distances globally, a global \gls{d_ave_criteria} is defined as follows:
\begin{equation}
    \label{eq:c5_distance_criteria}
    \displaystyle \gls{d_ave_criteria} = \left( \frac{ \int\limits_\Omega S_{\textrm T} T^2~d\Omega + \int\limits_\Omega S_{\textrm g} {\mathbf g}^2~d\Omega + \int\limits_\Omega S_{\textrm q} \mathbf q^2~d\Omega  }{\mu(\Omega)} \right)^{\nicefrac{1}{2}},
\end{equation}
where $\mu(\Omega)$ is the area of the domain $\Omega$ in a 2D case. 

A tolerance \gls{d_ave_tolerance} for the ratio between the average distance per element $\gls{d_ave}$ and the equivalent root mean squared measure \gls{d_ave_criteria} is introduced to determine which elements should not be taken into account for the adaptive $p$-refinement as follows: \textit{Do not increase the approximation order of the element if:}
\begin{equation}
    \label{eq:c5_order_not_refined}
    \gls{d_ave} > \gls{d_ave_tolerance} \; \gls{d_ave_criteria}.
\end{equation}
In a similar fashion, the standard deviation of the distance to the dataset \gls{d_std} can be compared to \gls{d_ave_criteria} to identify elements for $h$-refinement:
\begin{equation}
    \label{eq:c5_std_criteria}
    \gls{d_std} > \gls{d_std_tolerance} \; \gls{d_ave_criteria},
\end{equation}
where \gls{d_std_tolerance} is another tolerance threshold related to the standard deviation of the distance \eqref{eq:c5_distance_std}.

However, high standard deviation of the distance \eqref{eq:distance_dataset_dd_g} within an element does not mean that $h$-refining the element always yields improved results or provides more information about the problem uncertainty. For an element where all of the values at the integration points are too far from the dataset, i.e. have high distance \eqref{eq:distance_dataset_dd_g}, $h$-refinement does not provide any advantage. Therefore, an additional check is proposed which compares the values of the average and standard deviation in the element to evaluate if the refinement should take place:
\begin{equation}
    \label{eq:c5_std_high_2}
    \gls{d_std} > \gls{d_std_ave_tolerance} \; \gls{d_ave},
\end{equation}
where \gls{d_std_ave_tolerance} is another tolerance parameter. 

\subsection{Adaptive $hp^*$-refinement algorithm} \label{sec:adaptive_hp_refinement}

The error indicators \eqref{eq:error_indicator_grad}, \eqref{eq:error_indicator_div}, \eqref{eq:error_indicator_jump_sum} and estimator \eqref{eq:error_estimator}, are computed {\it a posteriori}, and highlight the elements with the highest FE approximation errors or the elements which are influenced by a less reliable part of the material dataset. Since the elements flagged by a high error estimator with high distance to the material dataset are not well informed of the material properties, the approximation order $p$ is not increased for these elements and the values of the error estimator on these elements should not be included in calculation of the average error estimator $\mu_{\textrm{avg}}$. Therefore, we introduce \textit{bounded} average error estimator $\hat{\mu}_{\textrm{avg}}$, which excludes the elements with high distance to the dataset. 

Let $\mathcal{I} := \left\{ e \in \{1, \dots, N\} \;\middle|\; d_\text{ave}^{e} \leq \gls{d_ave_tolerance} \right\}$, where $d_\text{ave}^{e}$ is the average distance to the dataset for the $e$-th element, be the set of indices of elements with field values sufficiently close to the dataset. The \textit{bounded} average error estimator is then given by:
\begin{equation}
\label{eq:bounded_average_error_estimator_set_based}
\hat{\mu}_{\textrm{avg}} =
\frac{1}{|\mathcal{I}|} \sum_{e \in \mathcal{I}} \mu_e,
\end{equation}
where $|\mathcal{I}|$ is the number of elements in the set $\mathcal{I}$.

Performing $p$-refinement\footnote{It is worth to note that a more robust $p$-refinement can be achieved if the algorithm does not allow two neighbouring elements to have a difference in approximation order more than two, see~\citep{kulikova_data-driven_2025}.} inside the domain where the finite element approximation errors are high and the distance to the material dataset is not too high, and $h$-refinement in elements adjacent to the boundaries and in elements with high standard deviation of the distance to the dataset, the adaptive refinement is performed as shown in \autoref{alg:c5_hp_refinement}.

\begin{algorithm} [tb]
    \caption{Adaptive $hp^*$-refinement driven by error estimator and distance to the material dataset}
    \label{alg:c5_hp_refinement}
    \begin{algorithmic}[1]
        \State Initialize with the lowest approximation order $p=1$ and the coarsest mesh.
        \Repeat
            \State Solve the problem \eqref{eq:mixed_variational}.
            \State Calculate the error indicators and estimator \eqref{eq:error_estimator}.
            \State Calculate the bounded average error estimator~\eqref{eq:bounded_average_error_estimator_set_based}.
            \State Calculate the distance to the material dataset at every integration point \eqref{eq:distance_dataset_dd_g}.
            \State Calculate the average distance \gls{d_ave} per element \eqref{eq:c5_distance_average}.
            \State Calculate the standard deviation of the distance \gls{d_std} per element \eqref{eq:c5_distance_std}.
            \State Calculate the comparative measurement \gls{d_ave_criteria} \eqref{eq:c5_distance_criteria}. 
            \For{each element $e$ in \{1, \dots, N\}}
                \If{$\mu_e > \hat{\mu}_{\textrm{avg}} \gls{err_est_order_tolerance}$ \textbf{and} $d_{ave}^e \le \gls{d_ave_tolerance} \gls{d_ave_criteria}$}
                    \State $p_e := p_e + 1$
                \EndIf
                \If{\{$\mu_e > \hat{\mu}_{\textrm{avg}} \gls{err_est_mesh_tolerance}$ \textbf{and} element is adjacent to boundary $\Gamma$ corner\}
                \textbf{or} 
                \{$d_{std}^e > \gls{d_std_tolerance} \gls{d_ave_criteria}$ \textbf{and} $d_{std}^e > \gls{d_std_ave_tolerance} d_{ave}^e$\} }
                    \State $h$-refine this element.
                \EndIf
            \EndFor
        \Until{a specified number of refinements is reached.}
    \end{algorithmic}
\end{algorithm}

\subsection{Demonstration of the adaptive refinement with error indicators}

The adaptive refinement is demonstrated on the Exp-Hat example with the \textit{tweaked} dataset, see~\autoref{fig:c5_hat_missing_err_est_full_ord_ref}(a). 
The refinement utilises \autoref{alg:c5_hp_refinement} with tolerances defined in \autoref{tab:tolerances}. These tolerances were chosen to achieve good convergence of adaptive iterations and at the same time highlight imperfections of the datasets considered in this paper. In the future work, a heuristic algorithm needs to be proposed to automatise the choice of these parameters. 

\begin{table} [tb]
    \centering
    \renewcommand{\arraystretch}{1.3} 
    \begin{tabular}{c|c|c|c|p{0.41\textwidth}}
        \textbf{Condition} & \textbf{Threshold} & \textbf{Value} & \textbf{Refinement} & \textbf{Description} \\ \hline
        $\mu_e > \mu_{\textrm{avg}} \gls{err_est_order_tolerance}$ & $\gls{err_est_order_tolerance}$ & 1.5 & $p$: $\uparrow$  & High error estimator: increase $p$ \\ 
        \hdashline
        $\gls{d_ave} > \gls{d_ave_tolerance} \; \gls{d_ave_criteria}$ & $\gls{d_ave_tolerance}$ & 4 & $p$: $\times$ & Too far from dataset: prevent refinement 
        \\ \hline
        $\gls{d_std} > \gls{d_std_tolerance} \; \gls{d_ave_criteria}$ & $\gls{d_std_tolerance}$ & 0.5 & $h$: $\uparrow$ & High standard deviation: refine mesh 
        \\ \hdashline
        $\gls{d_std} < \gls{d_std_ave_tolerance} \; \gls{d_ave}$ & $\gls{d_std_ave_tolerance}$ & 0.5 & $h$: $\times$ & Too far from dataset: prevent refinement \\ 
        \hline
        $\mu_e > \mu_{\textrm{avg}} \gls{err_est_mesh_tolerance}$ & $\gls{err_est_mesh_tolerance}$ & 4 & $h$: $\uparrow$ & Condition aimed at tackling flux singularities \\ 
    \end{tabular}
    \caption{
    Threshold value for adaptive $hp^*$-refinement algorithm applied to example in~\autoref{fig:c5_dist_ref}. Each condition determines whether to increase the polynomial order ($p$), refine the mesh ($h$), or suppress refinement based on error estimators and distance measures relative to the material dataset. The table is divided into three groups (separated by horizontal solid lines); $\uparrow$ means an increase in the  refinement, and $\times$ overrules $\uparrow$ within the group.
    }
    \label{tab:tolerances}
\end{table}

\begin{figure}[tb]
    \centering
    \includegraphics[width=0.95\textwidth]{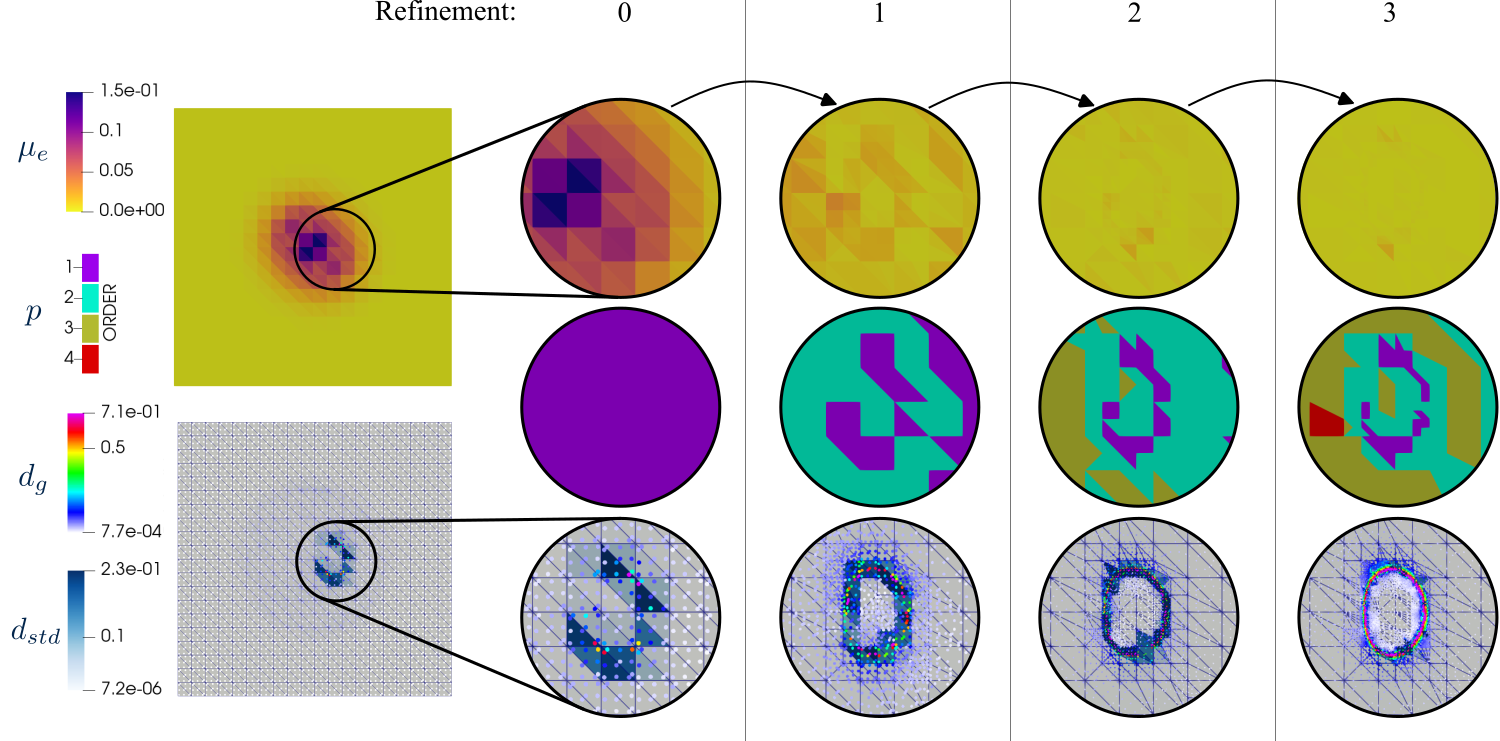}
    \caption{Snapshots of refinement iterations showing the error estimator (top row), approximation order (middle row), distance to the material dataset at each integration point and its standard deviation per element (bottom row). The refinement utilises \autoref{alg:c5_hp_refinement} with tolerances defined in \autoref{tab:tolerances}. }
    \label{fig:c5_dist_ref}
\end{figure}

\autoref{fig:c5_dist_ref} presents the snapshots of the refinement iterations, showing the distance to the material dataset at each integration point, its standard deviation per element, and the corresponding approximation order $p$. The algorithm is able to decrease the finite element approximation error where the knowledge of the material behaviour through the material dataset is high, while at the same time isolating the area with missing data by refining the mesh and keeping the approximation order low.

\subsection{Convergence w.r.t. exact solution}

\begin{figure} [tb]
    \centering
    \includegraphics[width=\textwidth]{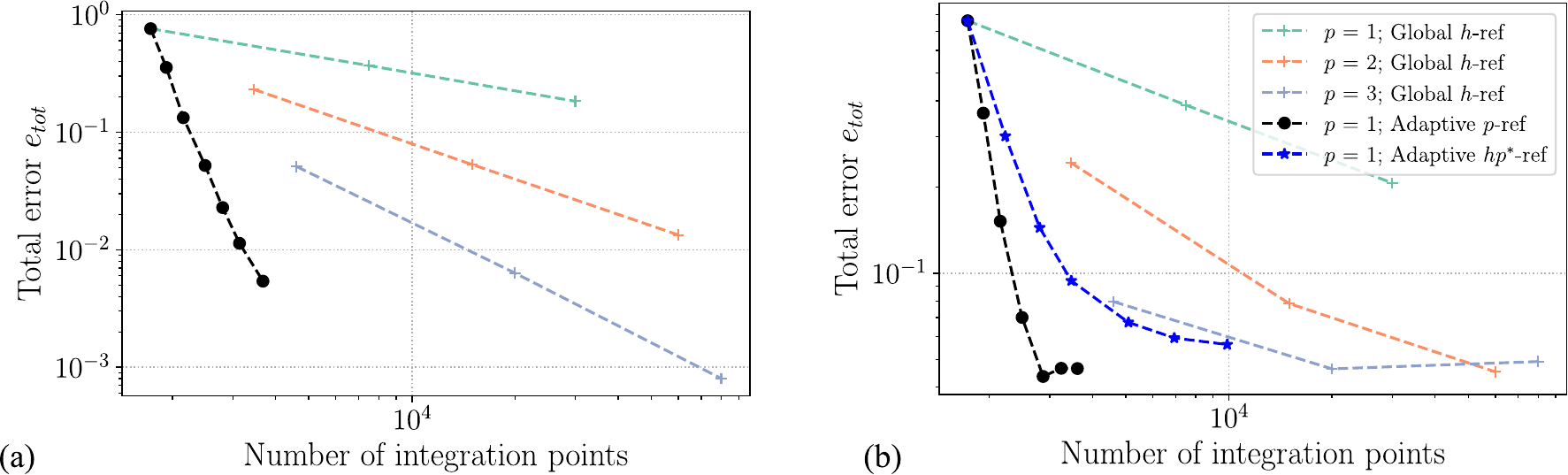}
    \caption{Convergence of the total global error $e_\text{tot}$ w.r.t. the exact solution using (a) ``saturated'' dataset and (b) \textit{tweaked} material dataset, see \autoref{fig:c5_hat_missing_err_est_full_ord_ref}(a). Note that adaptive $p$ and $hp^*$ refinements correspond to \autoref{alg:c5_p_refinement} and \autoref{alg:c5_hp_refinement}, respectively.}
    \label{fig:convergence_tot_error}
\end{figure}

To complete the discussion of adaptive refinement, a convergence study is performed on the Exp-Hat example with two material datasets introduced in the previous sections: ``saturated'' and \textit{tweaked}. 
Since the exact solution is known, a global total error $e_\text{tot}$ can be computed for the domain by summing all total errors per element \eqref{eq:total_error}. \autoref{fig:convergence_tot_error}(a) shows that when using the ``saturated'' material dataset, adaptive order refinement (\autoref{alg:c5_p_refinement}) provides convergence close to exponential. This is due to the nature of the problem, having highly localised flux, which is approximated better with higher orders of shape functions. 

On the contrary, \autoref{fig:convergence_tot_error}(b) shows that in case of the \textit{tweaked} dataset, the resulting total error converges to a plateau, which is caused by the missing data in the material dataset. The adaptive refinements prove to achieve the same accuracy as global refinement but with noticeably smaller number of integration points added to the problem. Note that the adaptive $p$-refinement proves to be the most efficient until reaching the plateau, however, the adaptive $hp^*$-refinement identifies and localises the problematic areas which are influenced by the missing material data, e.g. see \autoref{fig:c5_dist_ref}, for the cost of increased number of integration points. It should also be noted that this example uses conforming mesh refinement which leads to degradation of the resulting mesh and the results might improve with nonconforming mesh refinement~\citep{cerveny2019nonconforming}.

\section{Application for heat transfer in nuclear graphite} \label{sec:brick}

\begin{figure} [tb]
	\centering
	\includegraphics[width=\textwidth]{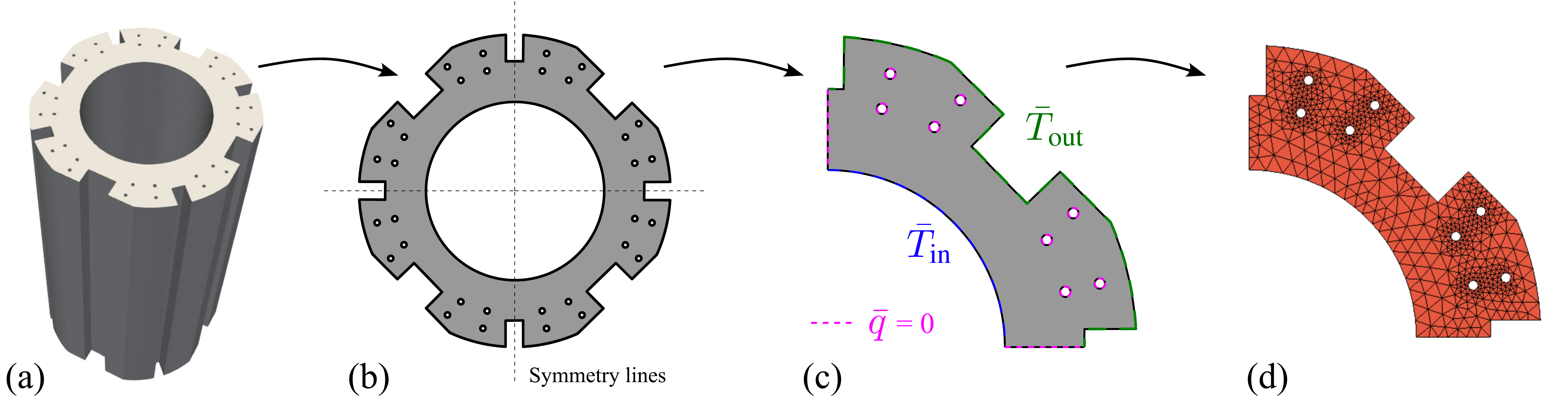}
	\caption{Graphite brick model: (a) 3D geometry, (b) 2D brick slice, (c) boundary conditions of $\bar T_{\textrm{in}} = 1000 ^\circ C$ and $\bar T_{\textrm{out}} = 500 ^\circ C$ and $\bar q = 0$ on the symmetry lines and the small inner holes, and (d) initial mesh}
	\label{fig:brick_setup}
\end{figure}

\begin{figure} [tb]
    \centering
    \includegraphics[width=\linewidth]{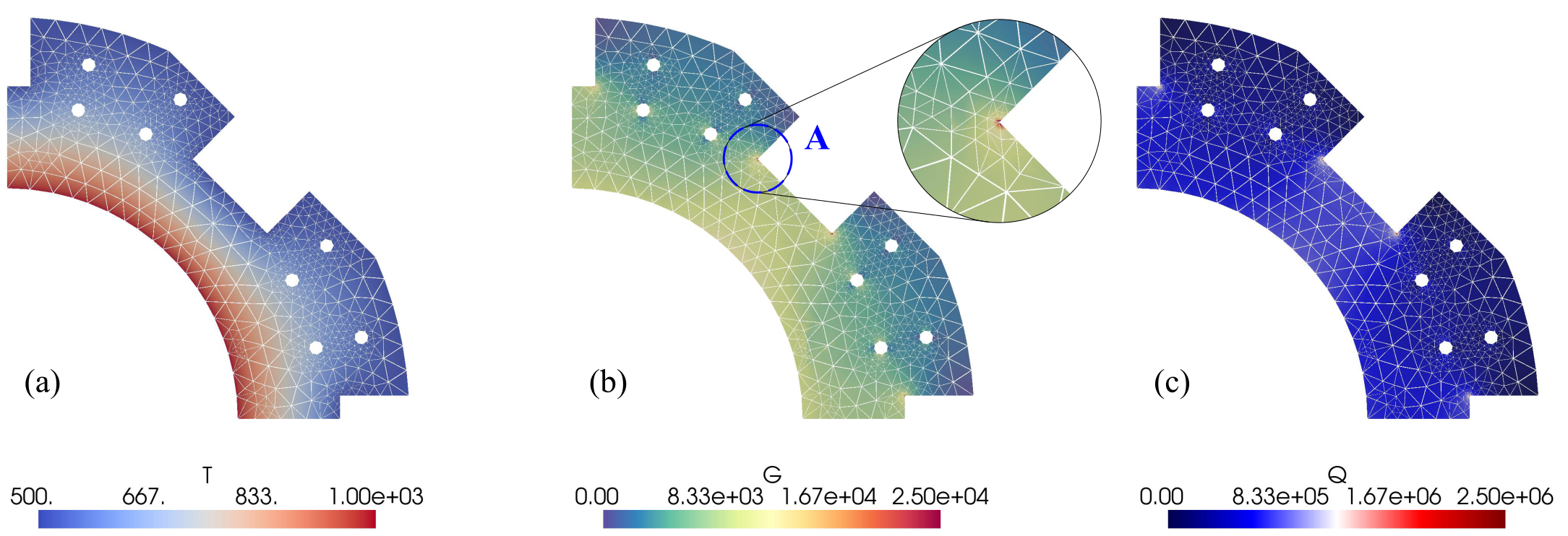}
    \caption{Results of the 2D brick slice example after $hp^*$-refinement: (a) temperature field $T$; (b) temperature gradient field $\mathbf g$; (c) flux field $\mathbf q$. The area of one of the singularities in the gradient and flux fields is marked by \textbf{A}.}
    \label{fig:brick_res}
\end{figure}

To demonstrate the application of the developed framework for an industrial example, we consider a 2D model of heat transfer in nuclear graphite brick used in Advanced gas-cooled reactors (AGR)~\citep{farrokhnia2022large,kelly1982graphite}, obtained from the 3D geometry by assuming no variations of the fields along the vertical axis of the brick, as shown in \autoref{fig:brick_setup}. Due to the symmetry of the problem and the boundary conditions applied, only a quarter of a horizontal brick slice is used in the numerical analysis. The inner boundary is kept at $1000 ^\circ C$ to simulate the exposure to the heat from the nuclear fuel housed in the cylindrical bore in the middle of the brick, while the outer boundary is kept at $500 ^\circ C$ representing the exposure to the carbon dioxide gas acting as a coolant when the reactor is operational. These boundary conditions are chosen for testing purposes only, as the temperatures for each of the graphite bricks vary throughout a nuclear reactor's core.  At the same time, it is assumed that there is no heat transfer through the small inner holes, which are filled with methane gas, a poor thermal conductor under reactor operating conditions. \autoref{fig:brick_res} shows the resulting fields of temperature $T$, temperature gradient $\mathbf g$, and flux $\mathbf q$ after the adaptive $hp^*$-refinement.

For testing purposes, the material dataset described in \autoref{fig:artificial_datasets}(b) and \autoref{sec:artificial_experiment} was used. This dataset is created through artificial experiments, and after the material datapoints were generated, a subset of points with $g_x^*$ in the range $[-9 \times 10^3 , -6 \times 10^3]$ $^\circ$C/m and $g_y^*$ in the range $[-4 \times 10^3 , -1 \times 10^3]$ $^\circ$C/m was removed. The artificial issue of missing data could represent equipment malfunction during testing, poor handling/preprocessing of the material datapoints, or simply not including the material states in the data gathering procedure. Furthermore, as real material datasets always include noise, for the purpose of this example, a white noise with a normal distribution and a standard deviation of $\sigma_\eta = 10^4$ [W/m$^2$] was added to the flux values in the $y$-direction for any points with temperature gradient magnitude less than $2 \times 10^{-3}$:
\begin{equation}
    \label{eq:c5_brick_noise}
    \begin{aligned}
        q_y^* :=
        \begin{cases}
            q_y^* + \eta & \textrm{if} \; |\mathbf g^*| < 2 \times 10^3 \\
            q_y^* & \textrm{otherwise},
        \end{cases}
    \end{aligned}
\end{equation}
where $\eta \sim  \mathcal{N}(0, \sigma_\eta)$. This scenario could represent a case where the measurement of small temperature gradient is difficult and/or inaccurate. The noise was only added to the flux values in the $y$-direction to evaluate the propagation of the noise to other directions.

\begin{figure}  [tb]
    \centering
    \includegraphics[width=\textwidth]{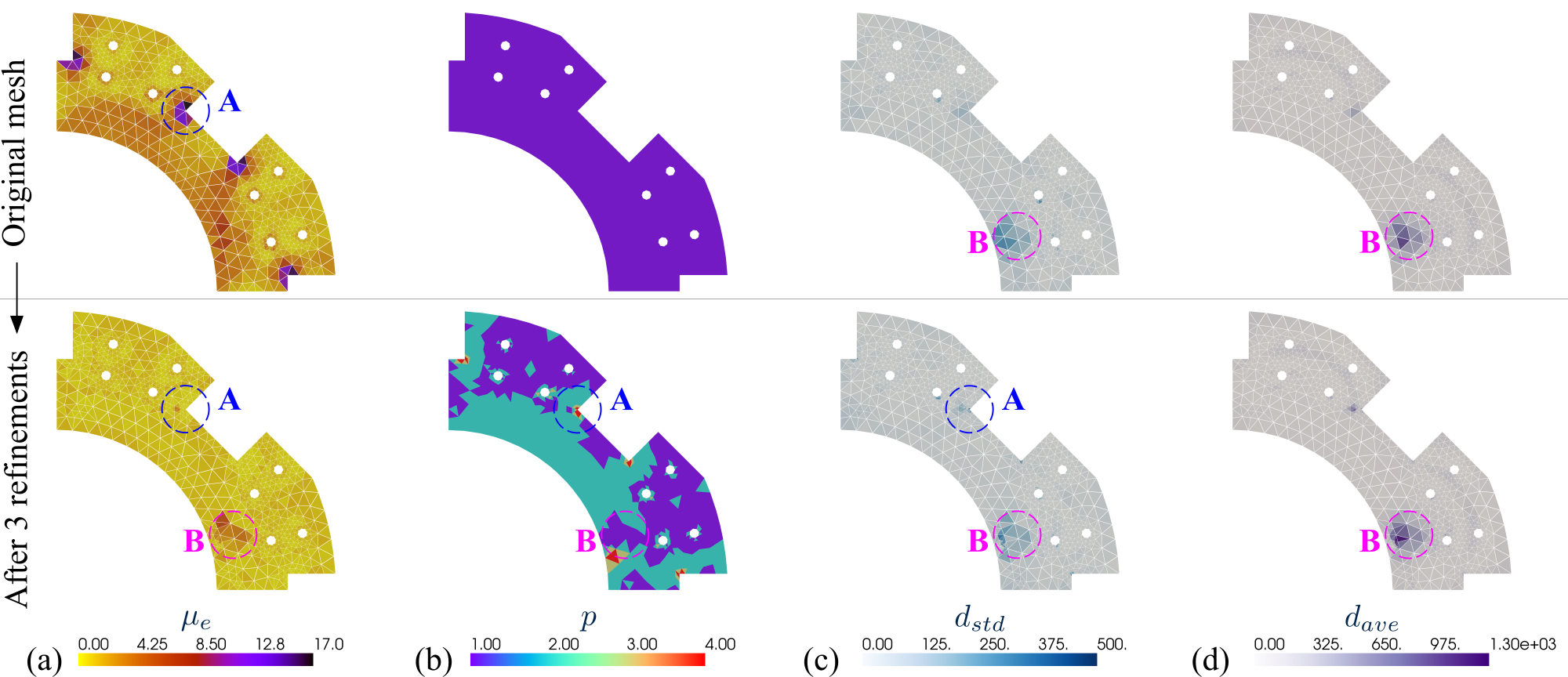}
    \caption{Comparison of the error indicators for the graphite brick example before and after using adaptive $hp^*$-refinement algorithm, showing snapshots of the (a) error estimator $\mu_e$, (b) approximation order $p$, (c) standard deviation and (d) average of the distance to the material dataset. Area \textbf{A} corresponds to one of the regions with flux singularity, and area \textbf{B} is the part of the domain where the field values are far from the dataset (since the dataset is missing data). The refinement utilises \autoref{alg:c5_hp_refinement} with tolerances defined in \autoref{tab:tolerances}.}
    \label{fig:brick_ref}
\end{figure}

The $hp^*$-refinement algorithm (\autoref{alg:c5_hp_refinement}) with tolerances as in \autoref{tab:tolerances} was used to refine the mesh and increase the approximation order for this problem. The adaptive refinement was stopped after 3 iterations and the result is compared to the solution without any refinement in \autoref{fig:brick_ref}. The approximation order was increased where the finite element approximation error was high within the domain and the distance to the material dataset was not \textit{too high}. Additionally, the mesh refinement handles well the area of high singularity of the heat flux $\mathbf q$ and temperature gradient $\mathbf g$, decreasing the finite element approximation error, see area \textbf{A} in \autoref{fig:brick_ref}(a). Furthermore, \autoref{alg:c5_hp_refinement} manages to decrease the standard deviation of the distance to the material dataset within elements where the average distance to the material dataset is not \textit{too high}, see area \textbf{B} in \autoref{fig:brick_ref}(c,~d). At the same time, the area where the distance to the material dataset is high is isolated and the approximation order is kept low there (see \autoref{fig:brick_ref}(b,~d), area \textbf{B}). As a result, the finite element approximation is not overfitted to imperfect data. 

\begin{figure}  [tb]
    \centering
    \includegraphics[width=\linewidth]{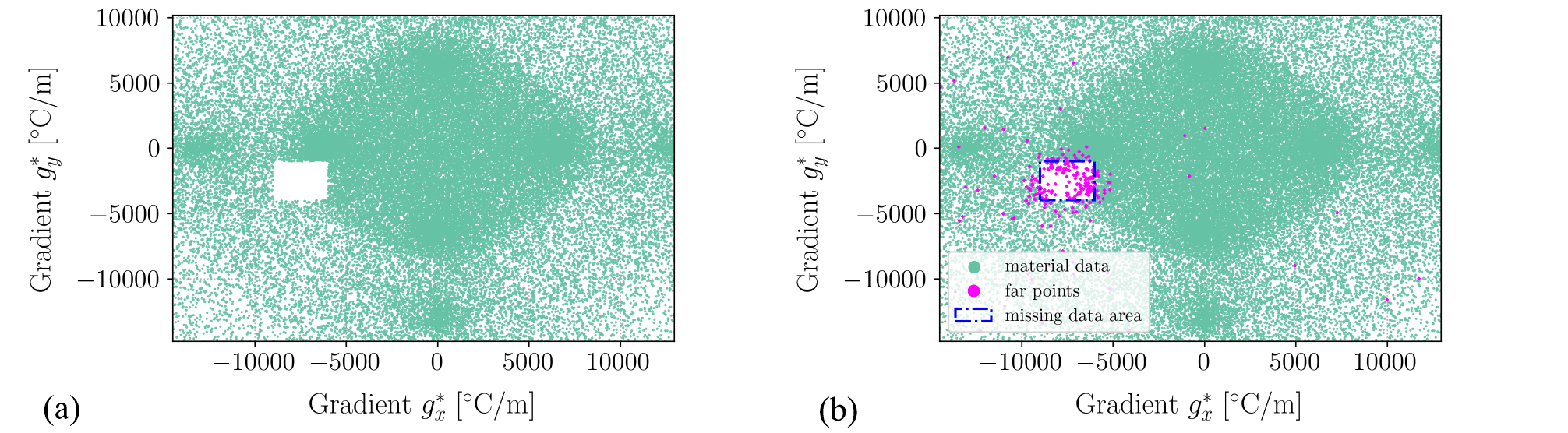}
    \caption{(a) Material dataset projection on $g_y^*$ and $g_x^*$ plane; (b) highlighting computed field values for integration points which are ``far'' from the material dataset according to \eqref{eq:c5_order_not_refined}.}
    \label{fig:mat_dataset_results_gradx_grady}
\end{figure}

The values of the calculated fields can be extracted w.r.t. the distance to the material dataset threshold \eqref{eq:c5_order_not_refined} defining which points are ``far'' from the material dataset upon convergence, see \autoref{fig:mat_dataset_results_gradx_grady}. Note that 72\% of ``far'' points obtained from the resulting field values are located inside and around the area where the material datapoints were removed from the dataset. This demonstrates the efficiency of the data-driven indicators proposed. However, some ``far'' points can be located elsewhere, e.g. in the regions where the dataset is too sparse for at least one of the fields.

In this section, we introduced several criteria to drive adaptive refinement, namely, finite element and distance error indicators. Application of the adaptive refinement algorithm reduces approximation errors and increases confidence in the predictions with an imperfect material dataset (e.g. missing data). However, the use of the material dataset enables multiple possible solutions to the problem which all satisfy the conservation laws and boundary conditions. Therefore, additional considerations are necessary to quantify the non-uniqueness of the solution caused by the dataset imperfections such as noise.

\section{Quantification of non-uniqueness} \label{sec:quantification_non-uniqueness}

The material dataset can contain noise or be missing data, and since there are multiple subsets of the material dataset which satisfy the conservation laws and boundary conditions, the solution is not unique. 
The certainty of the solution, therefore, depends on the quality of the material dataset w.r.t. the problem being solved. 
The uncertainty coming from the material dataset can be measured on the dataset level (preprocessing the material dataset), solution level (during the dataset search within the iterative process), or after the solution is obtained (postprocessing the results).
It is important to highlight that the non-uniqueness of the solution resulting from the DD formulation is not an issue and instead can be quantified and used to indicate the uncertainty of the solution.

At the same time, it is important to differentiate between the uncertainties/errors spanning from the finite element approximation and the uncertainties/errors coming from the material dataset used.
The adaptive refinement is therefore necessary to decrease the finite element approximation errors with minimal increase of the number of unknowns and integration points, which are closely related to the computational cost of the solution, see \autoref{sec:complexity} and \eqref{eq:complexity_dd}.

\begin{algorithm}[b]
    \caption{Markov chain Monte-Carlo simulations}
    \label{alg:monte_carlo}
    \begin{algorithmic}[1]
        \State Initialize the fields ($T, \mathbf g, \mathbf q$) with zero values.
        \State \label{alg:line:solve_refine}Solve the problem~\eqref{eq:mixed_variational} using the adaptive $hp^*$-refinement~\autoref{alg:c5_hp_refinement}.
        \Repeat
            \State Perturb the fields using \eqref{eq:pertutn_mcmc}.
            \State Set the perturbed values as initial fields.
            \State Solve the problem \eqref{eq:mixed_variational} using the refined mesh and order distribution obtained after step~\ref{alg:line:solve_refine}.
            \State Save the values of the fields. 
        \Until{Monte-Carlo criteria are satisfied (e.g. number of iterations performed)}
        \State Calculate the average and standard deviation using the saved values of the fields.
    \end{algorithmic}
\end{algorithm}

To address the uncertainties related to the material dataset, non-uniqueness quantification based on Markov chain Monte Carlo (MCMC) analysis is proposed, see \autoref{alg:monte_carlo}. In this algorithm, the solution of the problem using adaptive refinement, as outlined in~\autoref{alg:c5_hp_refinement}, serves as a starting point of the statistical analysis. The resulting fields are then perturbed in a random direction by a perturbation value, which is normally distributed as: 
\begin{equation}
    \label{eq:pertutn_mcmc}
    \gls{perturbation} \sim \mathcal{N} (0, \gls{std_perturbation} ),
\end{equation}
where \gls{std_perturbation} is a metaparameter corresponding to the standard deviation of the perturbation value. The perturbation metaparameter $\kappa$ can be chosen based on the average value of the field, the current local standard deviation of the field, the maximum value of the field, or the local or average distance to the material dataset, see~\citep{kulikova_data-driven_2025} for more details.

The fields requiring perturbation are the fields that are being solved for and are at the same time used in the search for the closest material data points, i.e. the flux field  $\mathbf q$, the gradient field $\mathbf g$ and the temperature field $T$. The perturbed fields are then used as initial fields for the next Monte-Carlo iteration of the analysis. 
The mesh and the approximation order are unchanged between the iterations and remain the same as at the end of the adaptive refinement. After each solution, the resulting fields are saved and the process is repeated a number of times until the Monte-Carlo criteria are satisfied. The stopping criteria can be set to the number of iterations or the difference between the average standard deviation of the fields from the previous and the current iteration. 

\begin{figure}  [tb]
    \centering
    \includegraphics[width=\textwidth]{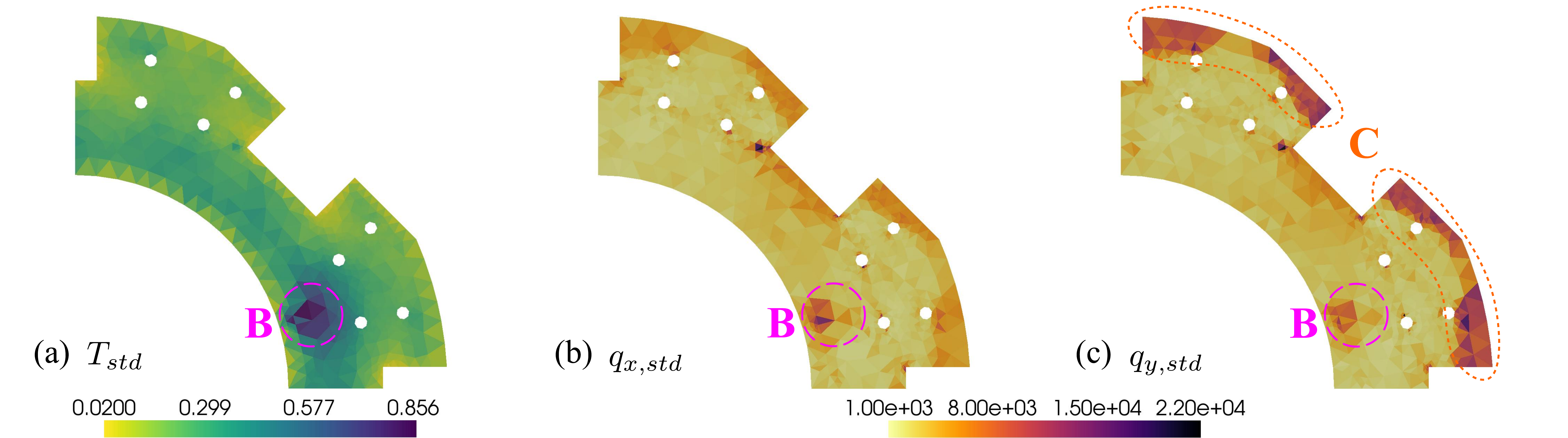}
    \caption{Standard deviation of the (a) temperature $T$ and (b-c) $x$ and $y$ directions of flux $\mathbf q$ fields after the adaptive refinement and Monte-Carlo simulations, showing the non-uniqueness of the solution w.r.t. the material dataset used. \textbf{B} is the area which corresponds to the part of the dataset which is missing points, and \textbf{C} is the area where the noise has been added to the $y$-direction of the flux values, see \eqref{eq:c5_brick_noise}.}
    \label{fig:brick_monte_carlo_results}
\end{figure}


Since the resulting fields are collected and saved after each iteration, the standard deviation of the fields can be calculated. The integration scheme used for calculating the fields is Gauss quadrature, with number of Gauss points chosen based on the approximation order of the fields. In contrast, the field values used for statistical analysis are evaluated and stored at regularly spaced locations within each element, determined using Newton-Cotes quadrature. These Newton-Cotes points are not used for integration; rather, they serve only as evaluation points to simplify the computation of uncertainty due to non-uniqueness.
In the example introduced, $\kappa = 10^4$ and the Monte-Carlo simulations are stopped after 100 iterations. 
The value of  $\kappa$ was chosen to be high enough to perturb the fields significantly, but not too high to affect the number of iterations required to converge the DD iterative process for each Monte-Carlo iteration, see~\citep{kulikova_data-driven_2025} for more details.

The results of the Monte-Carlo simulations are the average and the standard deviation of the fields evaluated on each regular evaluation point. The standard deviation for each field has the same units as the values of corresponding field; hence, its interpretation is straightforward. Furthermore, the standard deviation of each field is averaged per each element for the ease of visualisation, see \autoref{fig:brick_monte_carlo_results}.
In the areas not influenced by the missing data, the standard deviation of the flux field in the $y$-direction is higher than the standard deviation in the $x$-direction , see area \textbf{C} in \autoref{fig:brick_monte_carlo_results}(b,c), which is expected as the noise was added to the flux values only in the $y$-direction. Nevertheless, the results in the $x$-direction are still affected by the noise in the $y$-direction. 

\begin{figure} [tb]
    \centering
    \includegraphics[width=\linewidth]{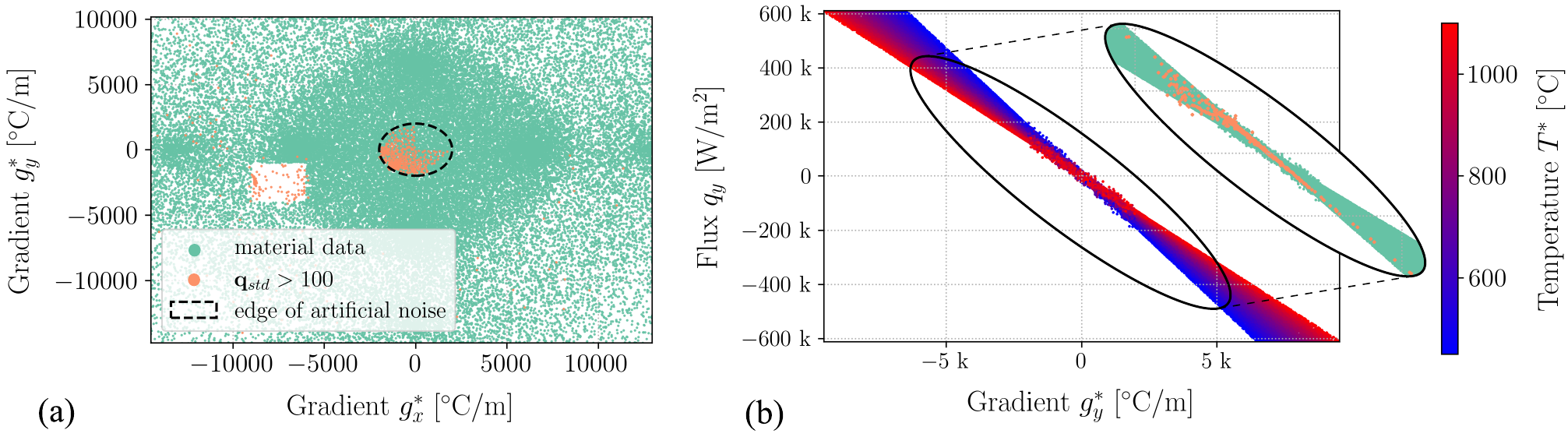}
    \caption{Points with high standard deviation of flux presented on the material dataset projections: (a) $g_x^*$ and $g_y^*$; (b) $g_y^*$ and $q_y^*$ which highlight the noisy points in the dataset.
    }
    \label{fig:high_std}
\end{figure}

The field values affected by the high standard deviations can be extracted from the analysis and plotted in the same projection as the material dataset, see \autoref{fig:high_std} for the points corresponding to flux standard deviation $\mathbf{q}_{std} > 100$ [W/m$^2$]. Note that 66\% of points are located inside the edge of the artificial noise, demonstrating the ability of~\autoref{alg:monte_carlo} to show the source of non-uniqueness. However, some of these points can be identified elsewhere, particularly in the region where material datapoints were removed (13\%), see area \textbf{B} in \autoref{fig:brick_monte_carlo_results}.

The standard deviation of the results should not, however, be the only indicator of the uncertainty of the solution as it only measures the non-uniqueness of the solution. 
Therefore, to interpret the results it is necessary to analyse the resulting fields (\autoref{fig:brick_res}), distance error indicators (Figures \ref{fig:brick_ref} and \ref{fig:mat_dataset_results_gradx_grady}) and standard deviation of the results (Figures \ref{fig:brick_monte_carlo_results} and \ref{fig:high_std}) together. 

\section{Conclusion}
\label{sec:conclusion}

This paper introduces a conservative data-driven finite element framework for solving heat transfer problems using material datasets directly in a simulation. The main novelty of this work is the proposed combination of the mixed finite element formulation and the data-driven approach. Mixed formulation uses the natural spaces for the unknown fields, allows for the jump of temperature between elements, and enforces the continuity of the normal flux component across any inner boundaries. Such a formulation provides naturally arising error indicators and estimator, which are used for adaptive refinement criteria.

The adaptive refinement algorithm was extended by calculating the variance of the distance to the material dataset within an element, which can be used together with the average distance to the material dataset per element to choose which elements should be refined.
The error indicators introduced through the mixed formulation were adapted to the weaker data-driven approach. The analysis showed that the standard error indicators highlight not only where the finite element approximation is poor, which can be improved by refinement, but also elements that use a problematic part of a material dataset. By using the proposed criteria for adaptive $hp$ refinement and choosing where $h$ and $p$ refinement takes place independently of each other, the algorithm is able to reduce the finite element approximation errors where the material dataset is relatively reliable and isolate areas where the material dataset is missing data.

Having used the adaptive refinement in the areas where it makes an impact on the accuracy, it is feasible to repeat the analysis multiple times with different initial field values to evaluate the uncertainty of the results. 
The final part of this work is devoted to quantifying the non-uniqueness of the resulting fields values using Markov chain Monte-Carlo simulations, evaluating the standard deviation of the results, and interpreting the non-uniqueness in terms of the material dataset used. This approach makes it possible to differentiate between the errors coming from the finite element approximation and uncertainty related to sparse, missing or noisy parts of material dataset.


The method is tested using a 2D model of a nuclear graphite brick, where the material dataset is created through artificial experiments. The adaptive refinement is able to target the flux singularities in the corners of the graphite brick and refine the elements accordingly, while at the same time isolating the area with missing data by refining the mesh and keeping the approximation order low. Finally, the uncertainty of the solution is quantified using Markov chain Monte Carlo simulations, which provide a measure of non-uniqueness of the solution.

The proposed method is a step towards a more flexible and efficient way of solving nonlinear heat transfer problems with material datasets, which can be used in various applications, such as nuclear engineering, materials science, and mechanical engineering. The method can be further extended to include more complex material datasets and to improve the adaptive refinement process.

The presented finite element technology for the data-driven approach offers a versatile tool for combining datasets from various laboratories and measurement techniques. The examples presented illustrate that, using such heterogeneous datasets, we can extract useful predictions and assess their quality. 

Furthermore, the implementation can be extended to non-conforming mesh refinement, which will aid the adaptive refinement accuracy by preventing deterioration in mesh quality, thereby minimizing approximation errors. To achieve this, a hybridized version of the mixed formulation can be implemented, where the $H(\text{div})$ space is ``broken'' introducing a hybridized field on the skeleton to enforce continuity of the tractions~\citep{boffi2013mixed}. This approach results in a sparser matrix with a block structure suitable for GPU computing~\citep{dobrev2019algebraic}. Such technology would be optimal, as it would offer rigorous error estimators, uncertainty quantification, and would be compatible with modern hardware architectures, providing an alternative to methods like deep neural networks and physics-informed neural networks (PINNs).

\section{Acknowledgements}

The first author would like to thank their PhD examiners for constructive comments. This work was supported by EDF Energy and UKRI EPSRC (ICASE EP/T517434/1). The views expressed in this document are those of the authors and not necessarily those of EDF Energy. 

\section{Declaration of generative AI and AI-assisted technologies in the writing process}

During the preparation of this work, the authors used Grammarly and Writefull (Overleaf) to check grammar and spelling mistakes. After using these tools/services, the authors reviewed and edited the content as needed and took full responsibility for the content of the published article.

\appendix

\newpage
\section{Dataset Creation}
\label{app:dataset_creation}
\setcounter{algorithm}{0}
\renewcommand{\thealgorithm}{A.\arabic{algorithm}}
\setcounter{figure}{0}

This appendix describes the details of the dataset creation process for the material datasets used in this work. Two approaches are introduced, termed \textit{regular} and \textit{artexp}. 

\subsection{Regularly spaced material dataset creation \label{app:regular_dataset}}

A synthetic dataset is generated by considering temperature gradients $\nabla T$ in the range $[-A, A]$, where the components $g_x = \partial T / \partial x$ and $g_y = \partial T / \partial y$ are sampled independently from a uniform distribution. The corresponding flux values are computed using the Fourier's law $\mathbf q = -k \nabla T$ with constant conductivity $k$, resulting in a grid of material data points, as illustrated in \autoref{fig:artificial_datasets}(a).

\autoref{alg:4D_dataset_regular} outlines the algorithm for constructing this regular dataset. It iterates over the specified range of temperature gradients, computes the flux components $q_x$ and $q_y$, and stores the resulting data points in the dataset $\mathcal{D}_{4D}$.

\begin{algorithm} 
    \caption{Creating a regular material dataset}
    \label{alg:4D_dataset_regular}
    \begin{algorithmic}[1]
        \State Initialise an empty dataset $\mathcal{D}_{4D}$
        \For{each temperature gradient $g_x$ in the range $[-A, A]$ with $count_G$ equally spaced intervals}
            \For{each temperature gradient $g_y$ in the range $[-A, A]$ with $count_G$ equally spaced intervals}
                \State Compute the flux components: $q_x = -k g_x$, $q_y = -k g_y$
                \State Add the point $(g_x, g_y, q_x, q_y)$ to the dataset $\mathcal{D}_{4D} = \{\mathbf{g}^*, \mathbf{q}^*\}$
            \EndFor
        \EndFor
    \end{algorithmic}
\end{algorithm}

\subsection{Artificial experiment material dataset creation \label{sec:artificial_experiment}} 

\begin{figure}  [tb]
    \centering
    \includegraphics[width=\textwidth]{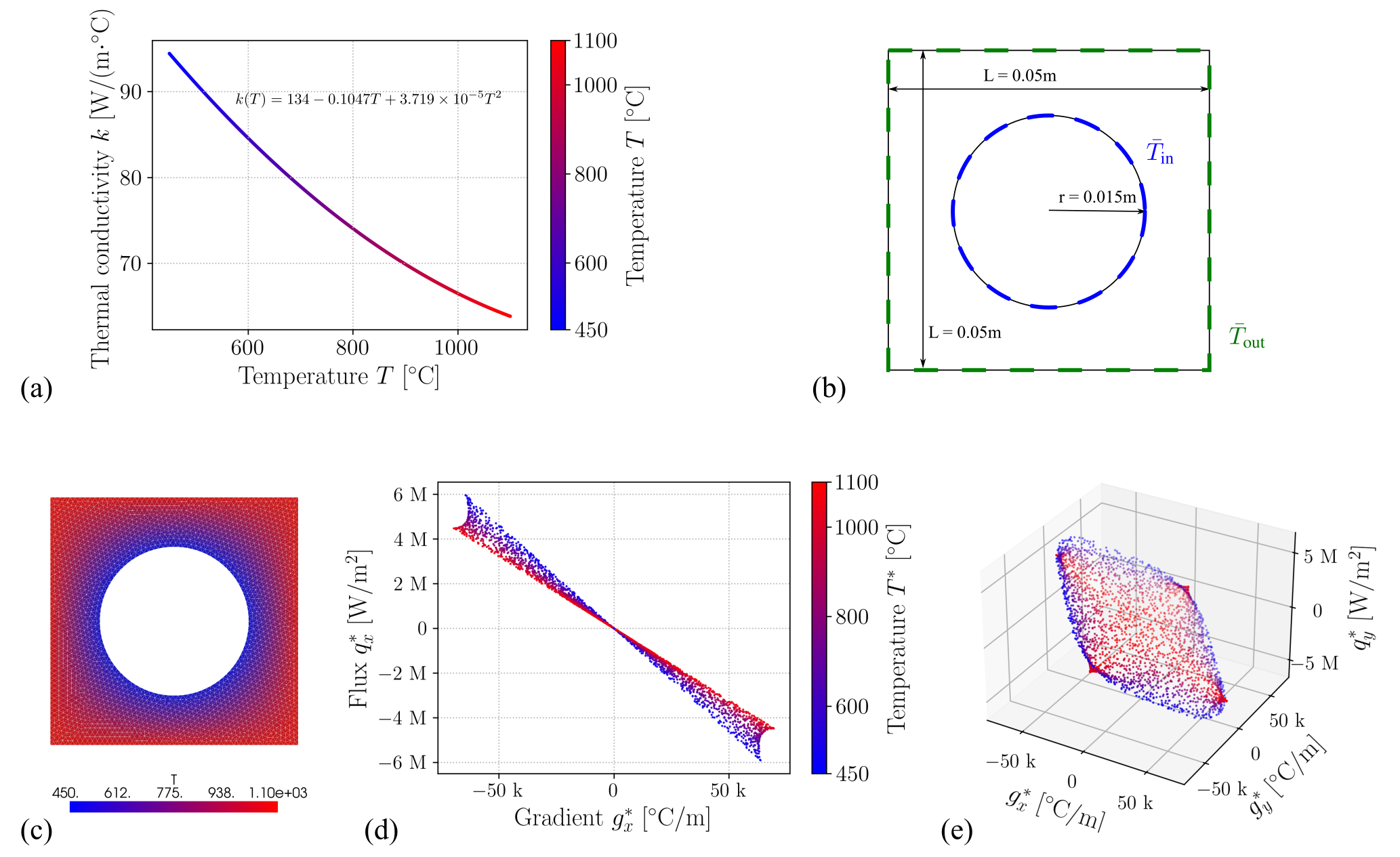}
    \captionsetup{justification=centering}
    \caption{Synthetic experiment example: (a) thermal conductivity w.r.t. temperature, (b) dimensions and boundary conditions of the experiment, (c) resulting temperature of $T_{\textrm{in}} = 450^\circ$C and $\bar T_{\textrm{out}} = 1100^\circ$C, (d-e) 2D and 3D views of resulting material dataset for (c).}
    \label{fig:artificial_experiment_1}
\end{figure}

The dataset for the graphite brick example is generated through a set of synthetic experiments using the material model for heat transfer in graphite~\eqref{eq:graphite_material_model}.
An illustration of such an experiment is shown in \autoref{fig:artificial_experiment_1}. The thermal conductivity as a function of temperature is shown in \autoref{fig:artificial_experiment_1}(a), and the corresponding geometry and boundary conditions are given in \autoref{fig:artificial_experiment_1}(b). The analysis employs hierarchical shape functions of approximation order $p = 2$ in a standard nonlinear finite element formulation derived from the conductivity model~\eqref{eq:graphite_material_model}; for further details, see~\cite{kulikova_data-driven_2025}. For one representative setup, the inner boundary temperature is fixed at $\bar{T}_{\textrm{in}} = 450^\circ$C and the outer boundary at $\bar{T}_{\textrm{out}} = 1100^\circ$C, yielding the temperature field shown in \autoref{fig:artificial_experiment_1}(c). The corresponding material dataset is visualised in 2D and 3D in \autoref{fig:artificial_experiment_1}(d-e), respectively. Additional experiments are performed by sweeping both $\bar{T}{\textrm{in}}$ and $\bar{T}{\textrm{out}}$ through 10 equally spaced values in the range $[400, 1100]$, and the full dataset is shown in \autoref{fig:artificial_datasets}(b).

\section{Finite element approximation for the data-driven formulation}

\subsection{Stronger formulation \label{app:fe_strong}}

The unknown fields in the variational problem~\eqref{eq:variational_dd} are approximated by the finite element basis functions as described in~\autoref{tab:DD_basis_functions}, see also~\autoref{tab:spaces_compared} for the definition of the corresponding functional spaces. Since \eqref{eq:variational_dd} holds for any $\delta T_{\gls{beta_H1}}$, $\delta q_{\gls{beta_L2}}$, and $\delta \lambda_{\gls{beta_H1}}$, the saddle point problem in a matrix form is written for each finite element as follows: 
\begin{equation}
    \label{eq:DD_matrix_form}
    \begin{bmatrix}
        \displaystyle\int\limits_{\Omega_e} S_{\textrm g} \nabla N_{\gls{beta_H1}} \cdot \nabla N_{\gls{alpha_H1}}~d\Omega & 0 & 0 \\
        0 & \displaystyle\int\limits_{\Omega_e} S_{\textrm q} I Q_{\gls{beta_L2}} Q_{\gls{alpha_L2}} ~d\Omega & \displaystyle\int\limits_{\Omega_e} \nabla L_{\gls{beta_H1}} Q_{\gls{alpha_L2}} ~d\Omega \\
		0 & \displaystyle\int\limits_{\Omega_e} \nabla L_{\gls{beta_H1}} Q_{\gls{alpha_L2}}~d\Omega & 0
    \end{bmatrix}
    \begin{bmatrix}
        T_{\gls{alpha_H1}} \\
        \mathbf q_{\gls{alpha_L2}} \\
		\lambda_{\gls{alpha_H1}}
    \end{bmatrix}
    =
    \begin{bmatrix}
        \displaystyle\int\limits_{\Omega_e} S_{\textrm g} \mathbf g^* \cdot \nabla N_{\gls{beta_H1}}~d\Omega \\
		\displaystyle\int\limits_{\Omega_e} S_{\textrm q} \mathbf q^*  Q_{\gls{beta_L2}}~d\Omega \\
        \displaystyle\int\limits_{\Gamma_{Te}} \bar q L_{\gls{beta_H1}} ~d\Gamma_T + \displaystyle\int\limits_{\Omega_e} s L_{\gls{beta_H1}}~d\Omega
    \end{bmatrix},
\end{equation}
where $T_{\gls{alpha_H1}}$, $\delta T_{\gls{beta_H1}}$, $\mathbf q_{\gls{alpha_L2}}$, $\delta \mathbf q_{\gls{beta_L2}}$, $\lambda_{\gls{alpha_H1}}$ and $\delta \lambda_{\gls{beta_H1}}$ are the coefficients of the approximation. are the unknown coefficients, $\Omega_e$ represents one finite element, and $\Gamma_{Te}$ corresponds to its boundary with the prescribed temperature.

\begin{table}[t]
	\centering
	\renewcommand{\arraystretch}{1.5} 
	\begin{tabular}{  >{\centering\arraybackslash}p{0.3\textwidth} 
		| >{\centering\arraybackslash}p{0.25\textwidth} : >{\centering\arraybackslash}p{0.25\textwidth}}
		& Trial function & Test function \\
		\hline
		Temperature & $T^h = T_{\gls{alpha_H1}} N_{\gls{alpha_H1}}$ & $\delta T^h = \delta T_{\gls{beta_H1}} N_{\gls{beta_H1}}$ \\
		\hdashline
		Flux & $\mathbf q^h = \mathbf q_{\gls{alpha_L2}} Q_{\gls{alpha_L2}}$ & $\delta \mathbf q^h = \delta \mathbf q_{\gls{beta_L2}} Q_{\gls{beta_L2}}$ \\
		\hdashline
		Lagrange multiplier & $\lambda^h = \lambda_{\gls{alpha_H1}} L_{\gls{alpha_H1}}$ & $\delta \lambda^h = \delta \lambda_{\gls{beta_H1}} L_{\gls{beta_H1}}$ \\
		\hline
		Gradient of temperature & $\nabla T^h = T_{\gls{alpha_H1}} \nabla N_{\gls{alpha_H1}}$ & $\nabla \delta T^h = \delta T_{\gls{beta_H1}} \nabla N_{\gls{beta_H1}}$ \\
		\hdashline
		Gradient of $\lambda$ & $\nabla \lambda^h = \lambda_{\gls{alpha_H1}} \nabla L_{\gls{alpha_H1}}$ & $\nabla \delta \lambda^h = \delta \lambda_{\gls{beta_H1}} \nabla L_{\gls{beta_H1}}$ \\
	\end{tabular}
	\caption{Finite element basis functions for the data-driven formulation of the diffusion problem. $T_{\gls{alpha_H1}}$, $\delta T_{\gls{beta_H1}}$, $\mathbf q_{\gls{alpha_L2}}$, $\delta \mathbf q_{\gls{beta_L2}}$, $\lambda_{\gls{alpha_H1}}$ and $\delta \lambda_{\gls{beta_H1}}$ are the coefficients of the approximation. $N_{\gls{alpha_H1}}$, $Q_{\gls{alpha_L2}}$ and $L_{\gls{alpha_H1}}$ are the scalar basis functions.}
	\label{tab:DD_basis_functions}
\end{table}

\subsection{Weaker formulation \label{app:fe_weak}}

The test and trial functions for the variational problem~\eqref{eq:mixed_variational} are approximated by the finite element basis functions as described in \autoref{tab:basis_functions_weaker_dd}. Note that both the gradient $\mathbf g$ and the flux $\mathbf q$ are vectorial fields, however, the components of gradient are approximated by scalar basis functions $G_{\gls{alpha_L2}}$, while the flux is approximated by a vectorial basis function $\mathbf Q_{\gls{alpha_Hdiv}}$. 
This is due to the fact that the gradient belongs to $\mathbf L^2(\Omega)$ space, while the flux belongs to $H(\text{div};\Omega)$ space, see~\autoref{tab:spaces_compared}.

The matrix form of the problem~\eqref{eq:mixed_variational} for each element reads:
\begin{equation}
    \label{eq:weaker_matrix_form}
    [\mathbf K_{\alpha \beta}] [\mathbf u_\alpha] = [ \mathbf F_\beta ] ,
\end{equation}
where $ [\mathbf u_\alpha]$ is the vector of unknown coefficients:
\begin{equation}
	\label{eq:weaker_unknown_vector}
	[\mathbf u_\alpha] = 
	\begin{bmatrix}
		T_{\gls{alpha_L2}} \\
		\mathbf g_{\gls{alpha_L2}} \\
		q_{\gls{alpha_Hdiv}} \\
		\lambda_{\gls{alpha_L2}} \\
		\boldsymbol \tau_{\gls{alpha_Hdiv}}
	\end{bmatrix},
\end{equation}
$[\mathbf K_{\alpha \beta}]$ is the ``diffusivity'' matrix:
\begin{equation}
	\label{eq:weaker_diffusivity_matrix}
	[\mathbf K_{\alpha \beta}] = 
	\begin{bmatrix}
		0 & 0 & \mathbf 0  & \mathbf 0 & \displaystyle\int\limits_{\Omega_e} N_{\gls{beta_L2}} \left(\nabla \cdot \mathbf M_{\gls{alpha_Hdiv}}\right)\,d\Omega\\
		\mathbf 0 & \displaystyle\int\limits_{\Omega_e} S_{\textrm{g}} G_{\gls{beta_L2}} G_{\gls{alpha_L2}}\,d\Omega & \mathbf 0 & \mathbf 0  &  \displaystyle\int\limits_{\Omega_e} G_{\gls{beta_L2}} \mathbf M_{\gls{alpha_Hdiv}}\,d\Omega\\
		\mathbf 0 & \mathbf 0 & \displaystyle\int\limits_{\Omega_e} S_{\textrm{q}} \left( \mathbf  Q_{\gls{beta_Hdiv}} \cdot \mathbf Q_{\gls{alpha_Hdiv}} \right)\,d\Omega & \displaystyle\int\limits_{\Omega_e} \left(\nabla \cdot \mathbf  Q_{\gls{beta_Hdiv}}\right) L_{\gls{alpha_L2}}\,d\Omega & \mathbf 0 \\
		\mathbf 0 & \mathbf 0 & \displaystyle\int\limits_{\Omega_e} L_{\gls{beta_L2}} \left( \nabla \cdot \mathbf Q_{\gls{alpha_Hdiv}} \right)\,d\Omega & \mathbf 0 & \mathbf 0 \\
		\displaystyle\int\limits_{\Omega_e} \left(\nabla \cdot \mathbf M_{\gls{beta_Hdiv}}\right) N_{\gls{alpha_L2}}\,d\Omega & \displaystyle\int\limits_{\Omega_e} \mathbf M_{\gls{beta_Hdiv}} G_{\gls{alpha_L2}}\,d\Omega & \mathbf 0 & \mathbf 0 & \mathbf 0 \\
	\end{bmatrix},
\end{equation}
and $[\mathbf F_\beta]$ is the vector of right-hand side terms:
\begin{equation}
	\label{eq:weaker_rhs_vector}
	[\mathbf F_\beta] = 
	\begin{bmatrix}
		\mathbf 0 \\
		\displaystyle\int\limits_{\Omega_e} S_{\textrm{g}} G_{\gls{beta_L2}} \mathbf g^*~d\Omega \\
		\displaystyle\int\limits_{\Omega_e} S_{\textrm{q}} \mathbf  Q_{\gls{beta_Hdiv}} \cdot \mathbf q^*~d\Omega \\
		\displaystyle\int\limits_{\Omega_e} L_{\gls{beta_L2}} s~d\Omega \\
		\displaystyle\int\limits_{\Gamma_{Te}} \left( \mathbf M_{\gls{beta_Hdiv}} \cdot \mathbf n \right)~ \bar T~ d \Gamma \\
	\end{bmatrix}.
\end{equation}

\begin{table}[t]
    \centering
    \renewcommand{\arraystretch}{1.5} 
    \begin{tabular}{  >{\centering\arraybackslash}p{0.25\textwidth} 
        | >{\centering\arraybackslash}p{0.25\textwidth} : >{\centering\arraybackslash}p{0.25\textwidth}}
     & Trial function & Test function \\
     \hline
    Temperature & $T^h = T_{\gls{alpha_L2}} N_{\gls{alpha_L2}}$ & $\delta T^h = \delta T_{\gls{beta_L2}} N_{\gls{beta_L2}}$ \\
    \hdashline
    Gradient & $\mathbf g^h = \mathbf g_{\gls{alpha_L2}} G_{\gls{alpha_L2}}$ & $\delta \mathbf g^h = \delta \mathbf g_{\gls{beta_L2}} G_{\gls{beta_L2}}$ \\
    \hdashline
    Flux & $\mathbf q^h = q_{\gls{alpha_Hdiv}} \mathbf Q_{\gls{alpha_Hdiv}}$ & $\delta \mathbf q^h = \delta  q_{\gls{beta_Hdiv}} \mathbf Q_{\gls{beta_Hdiv}}$ \\
	\hdashline
	Lagrange multiplier (scalar) & $\lambda^h = \lambda_{\gls{alpha_L2}} L_{\gls{alpha_L2}}$ & $\delta \lambda^h = \delta \lambda_{\gls{beta_L2}} L_{\gls{beta_L2}}$ \\ 
    \hdashline
    Lagrange multiplier (vectorial) & $\boldsymbol \tau^h = \tau_{\gls{alpha_Hdiv}} \mathbf M_{\gls{alpha_Hdiv}}$ & $\delta \boldsymbol \tau^h = \delta \tau_{\gls{beta_Hdiv}} \mathbf M_{\gls{beta_Hdiv}}$ \\
    \hline
    Divergence of flux & $\nabla \cdot \mathbf q^h = q_{\gls{alpha_Hdiv}} \left(\nabla \cdot \mathbf Q_{\gls{alpha_Hdiv}}\right)$ & $\nabla \cdot \delta \mathbf q^h = \delta  q_{\gls{beta_Hdiv}} \left(\nabla \cdot \mathbf  Q_{\gls{beta_Hdiv}}\right)$ \\
    \hdashline
    Divergence of $\boldsymbol \tau $ & $\nabla \cdot \boldsymbol \tau^h = \tau_{\gls{alpha_Hdiv}} \left(\nabla \cdot \mathbf M_{\gls{alpha_Hdiv}}\right)$ & $\nabla \cdot \delta \boldsymbol \tau^h = \delta \tau_{\gls{beta_Hdiv}} \left(\nabla \cdot \mathbf M_{\gls{beta_Hdiv}}\right)$ \\
    \end{tabular}
    \caption{Finite element basis functions for the weaker data-driven formulation of the diffusion problem.}
    \label{tab:basis_functions_weaker_dd}
\end{table}

\section{Finding the closest point on a line}  {\label{app:closest_point}}

The data search described by \eqref{eq:distance_dataset_dd} or \eqref{eq:distance_dataset_dd_g}, in case of ``saturated'' dataset, can be replaced by the search for the closest point on a curve corresponding to the constitutive equation $\mathbf q = -k \nabla T$. For simplicity, $\nabla T = \mathbf g$, and $(\cdot)^*$ refers to the value on the line in this section.

To find the closest point $\{ \mathbf g^*, \mathbf q^* \}_{4D}$ on the line 
\begin{equation}
	\label{eq:fourier_star}
	\mathbf q^* = -k \mathbf g^*
\end{equation}
to the field values $\{ \mathbf g, \mathbf q \}_{4D}$ at an integration point, the data search can be replaced by the following algorithm. 
First the distance of the point $\{ \mathbf g^*, \mathbf q^* \}_{4D}$ to the line \eqref{eq:fourier_star} is defined as:
\begin{equation}
    \label{eq:distance_line}
    \textrm{dist} \left(\{ \mathbf g, \mathbf q \}_{4D}, \mathbf q^* + k \mathbf g^* = 0 \right) = \frac{|\mathbf q + k \mathbf g|}{\sqrt{1+k^2}}  .
\end{equation}
Then the closest point on the line to the field values $\{ \mathbf g, \mathbf q \}_{4D}$ is found by:
\begin{equation}
	\label{eq:closest_point}
	\begin{aligned}
		\mathbf q^* &= \frac{k\left(k \mathbf q - \mathbf g\right)}{1+k^2} \\
		\mathbf g^* &= \frac{\left(-k \mathbf q + \mathbf g\right)}{1+k^2} ,
	\end{aligned}
\end{equation}
for a linear constitutive relationship. 




\bibliographystyle{elsarticle-num-names} 





\printglossary[type=symbols,title={List of Symbols}]

\end{document}